\documentclass[aps, prd, letterpaper, 12pt, nofootinbib, superscriptaddress, longbibliography, notitlepage]{revtex4-1}
\usepackage[utf8]{inputenc}
\usepackage{amsmath,amssymb,amsfonts}
\usepackage{mathrsfs}
\usepackage{color}
\usepackage{graphicx} 

\definecolor{hgreen}{rgb}{0,.7,0}
\definecolor{hred}{rgb}{.7,0,0}
\definecolor{hblue}{rgb}{0,0,.7}

\usepackage[colorlinks=true,
linkcolor=hblue,
citecolor=hgreen,
filecolor=hblue,
urlcolor=hred]{hyperref}

\allowdisplaybreaks

\begin{document}

\title{Probing Lepton Flavor Violation at \texorpdfstring{\\}{} Linear Electron-Positron Colliders}

\author{Wolfgang~Altmannshofer}
\email{waltmann@ucsc.edu}
\affiliation{Department of Physics, University of California Santa Cruz, and
Santa Cruz Institute for Particle Physics, 1156 High St., Santa Cruz, CA 95064, USA}

\author{Pankaj~Munbodh}
\email{pmunbodh@ucsc.edu}
\affiliation{Department of Physics, University of California Santa Cruz, and
Santa Cruz Institute for Particle Physics, 1156 High St., Santa Cruz, CA 95064, USA}

\begin{abstract}
The production of $\tau\mu$ pairs in electron-positron collisions offers a powerful probe of lepton flavor violation. In this work, we calculate the $e^+ e^- \to \tau \mu$ cross section within the framework of the Standard Model Effective Field Theory, allowing for arbitrary $e^+e^-$ beam polarizations. We then estimate the sensitivities of proposed future linear colliders, ILC and CLIC, to effective lepton flavor-violating interactions. The high center-of-mass energies achievable at these machines provide particularly strong sensitivity to four-fermion operators. Furthermore, the polarization of the $e^+e^-$ beams enables novel tests of the chirality structure of these interactions. We find that our projected sensitivities not only complement but in certain scenarios surpass those achievable with low-energy tau decay measurements at Belle~II.
\end{abstract}

\maketitle
\newpage
\tableofcontents
\newpage

\section{Introduction} \label{sec:intro}

In the Standard Model (SM), lepton flavor-violating (LFV) processes arise only at the loop level, and they are highly suppressed by the tiny ratio of neutrino masses to the weak scale~\cite{Marciano:1977wx, Petcov:1976ff, Lee:1977tib}. As a result, the predicted SM rates for LFV processes lie far below any conceivable experimental sensitivity. Hence, it is highly motivated to test LFV as it would be a clear indication of physics beyond the SM~\cite{Calibbi:2017uvl}.

Searches for LFV have traditionally focused on rare decays of muons and tau leptons. Experiments such as MEG~II, Mu3e, Mu2e, and COMET are pursuing muon decay channels and muon conversion~\cite{MEGII:2018kmf, COMET:2018auw, Bernstein:2019fyh, Mu3e:2020gyw, MEGII:2025gzr}, while Belle~II is expected to significantly improve the limits on LFV tau decays~\cite{Belle-II:2018jsg}. Complementary to these low-energy searches, high-energy colliders provide a different and powerful avenue to probe LFV interactions.

At high energies, one can search for lepton flavor-violating decays of heavy resonances such as $Z$ decays, $Z \to \ell \ell^\prime$~\cite{Delepine:2001di, Gutsche:2011bi, Davidson:2012wn, Calibbi:2021pyh}, Higgs decays, $H \to \ell \ell^\prime$~\cite{Davidson:2010xv, Blankenburg:2012ex, Harnik:2012pb, Altmannshofer:2015esa}, and also top decays, $t \to q \ell \ell^\prime$~\cite{Davidson:2015zza, Altmannshofer:2025lun} (see~\cite{Altmannshofer:2022fvz} for a review).
An alternative novel approach is to directly search for the non-resonant production of a lepton anti-lepton pair with different flavors in high energy collisions. This method takes advantage of the possible energy growth of the signal cross section compared to SM backgrounds that typically fall with the center-of-mass energy squared.

In this work, we focus on the LFV process $e^+ e^- \to \tau \mu$ at future electron-positron colliders.
We perform our analysis within the framework of the Standard Model Effective Field Theory (SMEFT)~\cite{Grzadkowski:2010es}, which allows us to parameterize possible new physics effects in a systematic and model-independent way as long as the new physics scale is sufficiently large compared to the center-of-mass energies of the collider, $\Lambda \gtrsim \sqrt{s}$.

In~\cite{Altmannshofer:2023tsa, Munbodh:2024shg}, we showed that the future circular colliders FCC-ee~\cite{Bernardi:2022hny, FCC:2025lpp} and CEPC~\cite{CEPCStudyGroup:2018ghi, CEPCStudyGroup:2023quu, Ai:2024nmn} have interesting sensitivities to SMEFT operators that contribute to the $e^+e^-\to \tau \mu$ process at the tree level. Utilizing the characteristic scaling of the $e^+e^-\to \tau \mu$ cross section with the center-of-mass energy we showed how one can disentangle the effects of different operator classes. The sensitivities to some combinations of the Wilson coefficients were found to be competitive with those from future $\tau$ decay searches at Belle~II \cite{Belle-II:2018jsg}. See also~\cite{Han:2010sa, Murakami:2014tna, Cho:2018mro, Cirigliano:2021img, Etesami:2021hex, Calibbi:2022ddo, Barik:2023bgx, Lichtenstein:2023iut, Liu:2024gui, Jahedi:2024kvi, De:2024foq} for related studies.

In this work, we extend our analysis to linear $e^+ e^-$ colliders, such as the International Linear Collider (ILC)~\cite{ILC:2013jhg, Bambade:2019fyw, ILCInternationalDevelopmentTeam:2022izu} and the Compact Linear Collider (CLIC)~\cite{Linssen:2012hp, CLIC:2018fvx, Brunner:2022usy}. These machines offer high center-of-mass energies, large integrated luminosities, and the ability to polarize the electron and positron beams, giving them unique capabilities to test new physics in $e^+e^-\to \tau \mu$. 

In particular, we exploit the fact that the four-fermion operator contributions to the $e^+e^-\to \tau \mu$ cross section scale linearly with the squared center-of-mass energy, $s$.
Since the linear colliders are proposed to run at much higher $\sqrt{s}$ than circular colliders, we gain exceptional sensitivity to the four-fermion contact interactions, in many cases exceeding the projected Belle~II sensitivity of low energy tau decays. An additional advantage of linear colliders is their capacity for polarized electron and positron beams. With unpolarized beams, operators with similar tensorial structure enter the cross section in identical linear combinations, making it difficult to disentangle their individual contributions. Polarization offers a crucial handle to probe the chirality structure of the LFV operators, allowing us to meaningfully separate and constrain different operator classes.

The paper is organized as follows. In section~\ref{sec:crosssection}, we compute the $e^+ e^- \to \tau \mu$ cross section including all relevant SMEFT operators and arbitrary beam polarizations. In section~\ref{sec:sensitivities}, we determine the sensitivities of the linear colliders ILC and CLIC to the $e^+ e^- \to \tau \mu$ process. To distinguish the $e^+ e^- \to \tau \mu$ signal from $e^+e^- \to \tau^+\tau^-$ backgrounds we focus on the muon momentum as the discriminating kinematic variable. We discuss in detail the muon momentum distributions of signal and background, including detector resolution, beam energy spread, and initial state radiation. In section~\ref{sec:results}, we present our results for the ILC and CLIC sensitivity to the lepton flavor-violating SMEFT operators and compare them to Belle~II projections and our sensitivity estimates for future circular $e^+e^-$ colliders from~\cite{Altmannshofer:2023tsa}. Section~\ref{sec:conclusions} contains our conclusions.

\section{Signal cross-section} \label{sec:crosssection}

There are three main classes of operators that contribute to the process $e^+e^- \to \tau\mu$ at tree-level in the SMEFT: dipole operators, Higgs-lepton operators, and four-fermion operators. For a detailed description of the setup, we refer the reader to~\cite{Altmannshofer:2023tsa} (see also \cite{Crivellin:2013hpa, Celis:2014asa, Fernandez-Martinez:2024bxg}). Here, we only recall the convenient basis that we chose to work in. It consists of flavor-violating dipole operators formulated in terms of the photon and $Z$ boson field strengths $F_{\mu \nu}$ and $Z_{\mu \nu}$
\begin{align}
    (C_\gamma^{LR})_{\mu\tau} & \frac{1}{\sqrt{2}} \frac{v}{\Lambda^2} (\bar \mu \sigma^{\alpha\beta} P_R\tau) F_{\alpha\beta} ~, \quad & (C_Z^{LR})_{\mu\tau} & \frac{1}{\sqrt{2}} \frac{v}{\Lambda^2} (\bar \mu \sigma^{\alpha\beta} P_R\tau) Z_{\alpha\beta} ~, \\
    (C_\gamma^{RL})_{\mu\tau} & \frac{1}{\sqrt{2}} \frac{v}{\Lambda^2} (\bar \mu \sigma^{\alpha\beta} P_L \tau) F_{\alpha\beta} ~, \quad & (C_Z^{RL})_{\mu\tau} & \frac{1}{\sqrt{2}} \frac{v}{\Lambda^2} (\bar \mu \sigma^{\alpha\beta} P_L \tau) Z_{\alpha\beta} ~,
\end{align}
flavor-violating $Z$ couplings that are provided by the Higgs-lepton operators
\begin{align}
    (C_Z^{LL})_{\mu\tau} & \frac{v^2}{2\Lambda^2} (\bar \mu \gamma^{\alpha} P_L \tau) \frac{g}{c_W} Z_{\alpha} ~, \quad & (C_Z^{RR})_{\mu\tau} & \frac{v^2}{2\Lambda^2} (\bar \mu \gamma^{\alpha} P_R \tau) \frac{g}{c_W} Z_{\alpha} ~,
\end{align}
and the following flavor-violating four-fermion operators
\begin{align}
    (C_V^{LL})_{\mu\tau} & \frac{1}{\Lambda^2} (\bar e \gamma_\alpha P_L e)(\bar \mu \gamma^{\alpha} P_L \tau) ~, \quad & (C_V^{RR})_{\mu\tau} & \frac{1}{\Lambda^2} (\bar e \gamma_\alpha P_R e)(\bar \mu \gamma^{\alpha} P_R \tau) ~, \\
    (C_V^{LR})_{\mu\tau} & \frac{1}{\Lambda^2} (\bar e \gamma_\alpha P_L e)(\bar \mu \gamma^{\alpha} P_R \tau) ~, \quad & (C_V^{RL})_{\mu\tau} & \frac{1}{\Lambda^2} (\bar e \gamma_\alpha P_R e)(\bar \mu \gamma^{\alpha} P_L \tau) ~, \\
    (C_S^{LR})_{\mu\tau} & \frac{1}{\Lambda^2} (\bar e P_L e)(\bar \mu P_R \tau) ~, \quad & (C_S^{RL})_{\mu\tau} & \frac{1}{\Lambda^2} (\bar e P_R e)(\bar \mu P_L \tau) ~.
\end{align}
In the above expressions, $v \simeq 246$\,GeV is the Higgs vacuum expectation value (vev), $g$ is the weak gauge coupling, and $c_W = \cos\theta_W$ with the weak mixing angle $\theta_W$. Below we will also make use of the short-hand notation $s_W = \sin\theta_W$.
The Wilson coefficients of the above operators are linear combinations of SMEFT Wilson coefficients
\begin{align}
\label{eq:CgammaLR}
(C_{\gamma}^{LR})_{\mu \tau} &= c_W (C_{eB})_{\mu\tau} - s_W (C_{eW})_{\mu \tau} ~, \quad & (C_Z^{LR})_{\mu \tau} &= - c_W (C_{eW})_{\mu \tau} - s_W (C_{eB})_{\mu \tau} ~, \\
(C_{\gamma}^{RL})_{\mu\tau} &= c_W (C_{eB})^*_{\tau\mu} - s_W (C_{eW})^*_{\tau\mu} ~, \quad & (C_Z^{RL})_{\mu \tau} &= - c_W (C_{eW})^*_{\tau\mu} - s_W (C_{eB})^*_{\tau\mu} ~, \\
(C_Z^{LL})_{\mu \tau} &= (C_{\varphi \ell}^{(1)})_{\mu \tau} + (C_{\varphi \ell}^{(3)})_{\mu \tau}~, \quad &  (C_Z^{RR})_{\mu \tau} &= (C_{\varphi e})_{\mu \tau} ~, \\
(C_V^{LR})_{\mu \tau} &= (C_{\ell e})_{ee\mu\tau} ~, \quad & (C_V^{RL})_{\mu \tau} &= (C_{\ell e})_{\mu\tau ee} ~,  \\
(C_S^{LR})_{\mu \tau} &= - 2(C_{\ell e})_{\mu ee \tau} ~, \quad & (C_S^{RL})_{\mu \tau} &= - 2 (C_{\ell e})_{e \tau \mu e} ~,
\end{align}
\begin{align}
(C_V^{LL})_{\mu \tau} &= (C_{\ell\ell})_{ee\mu\tau} + (C_{\ell\ell})_{\mu\tau ee} + (C_{\ell\ell})_{e\tau \mu e} + (C_{\ell\ell})_{\mu e e\tau} ~, \\
\label{eq:CVRR}
(C_V^{RR})_{\mu \tau} &= (C_{ee})_{ee\mu\tau} + (C_{ee})_{\mu\tau ee} + (C_{ee})_{e\tau\mu e} + (C_{ee})_{\mu ee \tau} ~.
\end{align}

Linear electron-positron colliders typically use polarized beams, with the polarizations of the electron beam, $P_-$, and the positron beam, $P_+$, defined as (see e.g.~\cite{ILCInternationalDevelopmentTeam:2022izu})
\begin{equation}
P_- = \frac{N_{e^-_R} - N_{e^-_L}}{N_{e^-_R} + N_{e^-_L}} ~,\quad P_+ = \frac{N_{e^+_R} - N_{e^+_L}}{N_{e^+_R} + N_{e^+_L}} ~,
\end{equation}
where $N_{e^\pm_{L,R}}$ corresponds to the number of left-handed and right-handed electrons or positrons in the beams.

With these definitions, the differential cross section for $\mu^- \tau^+$ production in electron positron collisions can be written as
\begin{multline} \label{eq:dsigma}
 \frac{d\sigma (e^+ e^- \to \mu^- \tau^+)}{d\cos\theta} = \frac{1}{4} \Bigg[ (1 + P_+)(1 + P_-) \frac{d\sigma (e^+_R e^-_R \to \mu^- \tau^+)}{d\cos\theta} + (1 - P_+)(1 - P_-) \frac{d\sigma (e^+_L e^-_L \to \mu^- \tau^+)}{d\cos\theta} \\ + (1 + P_+)(1 - P_-) \frac{d\sigma (e^+_R e^-_L \to \mu^- \tau^+)}{d\cos\theta} + (1 - P_+)(1 + P_-) \frac{d\sigma (e^+_L e^-_R \to \mu^- \tau^+)}{d\cos\theta}  \Bigg] ~, 
\end{multline}
where $\theta$ is defined as the angle between the incoming electron and the outgoing muon (or, equivalently, the incoming positron and the outgoing anti-muon).
A completely analogous expression holds for the $\mu^+ \tau^-$ final state.

We find that the differential cross sections for specific electron and positron chiralities can be written as
\begin{equation}
\label{eq:cross_start}
    \frac{d\sigma (e^+_L e^-_R \to \mu^- \tau^+)}{d\cos\theta} = \frac{m_Z^2}{32\pi \Lambda^4} \Big[ I_0^{e^+_L e^-_R}(s) (1+\cos^2\theta) + I_1^{e^+_L e^-_R}(s) \cos \theta + I_2^{e^+_L e^-_R}(s) \sin^2\theta \Big] ~, 
\end{equation}
\begin{equation}
    \frac{d\sigma (e^+_R e^-_R \to \mu^- \tau^+)}{d\cos\theta} = \frac{m_Z^2}{32\pi \Lambda^4} I^{e^+_R e^-_R}(s) ~, 
\end{equation}

\begin{equation}
     \frac{d\sigma (e^+_L e^-_R \to \mu^+ \tau^-)}{d\cos\theta} = \frac{m_Z^2}{32\pi \Lambda^4} \Big[ \bar I_0^{e^+_L e^-_R}(s) (1+\cos^2\theta) + \bar I_1^{e^+_L e^-_R}(s) \cos\theta + \bar I_2^{e^+_L e^-_R}(s) \sin^2\theta \Big] ~,
\end{equation}
\begin{equation} 
     \label{eq:cross_end}
     \frac{d\sigma (e^+_R e^-_R \to \mu^+ \tau^-)}{d\cos\theta} = \frac{m_Z^2}{32\pi \Lambda^4} \bar I^{e^+_R e^-_R}(s) ~.
\end{equation}
Completely analogous expressions hold when exchanging the polarizations $R \leftrightarrow L$.

The total integrated cross section can then be written as
\begin{multline} \label{eq:sigma_tot}
\sigma(e^+e^- \to \tau\mu) = \int d\cos\theta \left( \frac{d\sigma (e^+ e^- \to \tau^+ \mu^-)}{d\cos\theta} + \frac{d\sigma (e^+ e^- \to \tau^- \mu^+)}{d\cos\theta} \right) \\
= \frac{m_Z^2}{64 \pi \Lambda^4} \bigg[ (1 + P_+)(1 + P_-) \Big( I^{e^+_R e^-_R}(s) + \bar I^{e^+_R e^-_R}(s) \Big) + (1 - P_+)(1 - P_-) \Big( I^{e^+_L e^-_L}(s) + \bar I^{e^+_L e^-_L}(s) \Big) \\ + \frac{2}{3} (1 + P_+)(1 - P_-) \Big( 2 I_0^{e^+_R e^-_L}(s) + 2 \bar I_0^{e^+_R e^-_L}(s) + I_2^{e^+_R e^-_L}(s) + \bar I_2^{e^+_R e^-_L}(s)\Big) \\ + \frac{2}{3}(1 - P_+)(1 + P_-)\Big( 2 I_0^{e^+_L e^-_R}(s) + 2 \bar I_0^{e^+_L e^-_R}(s) + I_2^{e^+_L e^-_R}(s) + \bar I_2^{e^+_L e^-_R}(s)\Big) \bigg] ~.
\end{multline}

For the coefficient functions $I_i(s)$, we find the following results
\begin{multline}
I_0^{e^+_L e^-_R}(s) = \frac{s}{m_Z^2} \Big( |C_V^{RL}|^2 + |C_V^{RR}|^2 \Big) + \frac{4 s_W^2 s^2}{(s-m_Z^2)^2 + \Gamma_Z^2 m_Z^2} \Bigg[ \frac{m_Z^2}{s} \Big( |C_Z^{LL}|^2 + |C_Z^{RR}|^2 \Big) s_W^2 \\  - \left( 1 - \frac{m_Z^2}{s} \right) \text{Re} \left( C_V^{RL} C_Z^{LL\star} + C_V^{RR} C_Z^{RR\star}\right)  + \frac{\Gamma_Z m_Z}{s} \text{Im} \left( C_V^{RL}C_Z^{LL\star} + C_V^{RR}C_Z^{RR\star} \right) \Bigg] ~,
\end{multline}

\begin{multline}
I_1^{e^+_L e^-_R}(s) = \frac{2s}{m_Z^2} \Big( |C_V^{RR}|^2 - |C_V^{RL}|^2 \Big) +  \frac{8 s_W^2 s^2}{(s-m_Z^2)^2+\Gamma_Z^2 m_Z^2} \Bigg[ \frac{m_Z^2}{s} \Big( |C_Z^{RR}|^2 - |C_Z^{LL}|^2 \Big) s_W^2 \\ - \left( 1 - \frac{m_Z^2}{s} \right) \text{Re} \left ( C_V^{RR} C_Z^{RR\star}-C_V^{RL} C_Z^{LL\star}\right) + \frac{\Gamma_Z m_Z}{s} \text{Im} \left( C_V^{RR} C_Z^{RR\star}- C_V^{RL} C_Z^{LL\star} \right) \Bigg] ~,
\end{multline}

\begin{multline}
I_2^{e^+_L e^-_R}(s) = 8 \Big( |C_\gamma^{LR}|^2 + |C_\gamma^{RL}|^2 \Big) s^2_W c^2_W +  \frac{16 s_W^3 c_W s^2}{(s-m_Z^2)^2 + \Gamma_Z^2 m_Z^2 } \Bigg[ \Big( |C_Z^{LR}|^2 + |C_Z^{RL}|^2 \Big) \frac{s_W}{2c_W} \\ - \left( 1 - \frac{m_Z^2}{s} \right) \text{Re} \left( C_\gamma^{LR} C_Z^{LR\star} + C_\gamma^{RL} C_Z^{RL\star}\right) + \frac{\Gamma_Z m_Z}{s} \text{Im} \left(C_\gamma^{LR} C_Z^{LR\star} + C_\gamma^{RL} C_Z^{RL\star} \right)  \Bigg] ~,
\end{multline}

\begin{multline}
I_0^{e^+_R e^-_L}(s) = \frac{s}{m_Z^2} \Big( |C_V^{LL}|^2 + |C_V^{LR}|^2 \Big) + \frac{2 ( 1 - 2 s_W^2) s^2}{(s-m_Z^2)^2 + \Gamma_Z^2 m_Z^2} \Bigg[  \frac{m_Z^2}{s} \Big( |C_Z^{LL}|^2 + |C_Z^{RR}|^2 \Big) \left( \frac{1}{2}-s_W^2\right) \\ + \left( 1 - \frac{m_Z^2}{s} \right) \text{Re} \left( C_V^{LL} C_Z^{LL\star} + C_V^{LR} C_Z^{RR\star} \right) - \frac{\Gamma_Z m_Z}{s} \text{Im} \left( C_V^{LL} C_Z^{LL\star} + C_V^{LR} C_Z^{RR\star}\right) \Bigg ] ~,
\end{multline}

\begin{multline}
I_1^{e^+_R e^-_L}(s) = \frac{2s}{m_Z^2} \Big( |C_V^{LL}|^2 - |C_V^{LR}|^2 \Big) + \frac{4 (1- 2s_W^2) s^2}{(s-m_Z^2)^2 + \Gamma_Z^2 m_Z^2} \Bigg[ \frac{m_Z^2}{s} \Big( |C_Z^{LL}|^2 - |C_Z^{RR}|^2 \Big)\left( \frac{1}{2}-s_W^2\right) \\ + \left( 1 - \frac{m_Z^2}{s} \right) \text{Re} \left( C_V^{LL} C_Z^{LL\star} - C_V^{LR} C_Z^{RR\star} \right) - \frac{\Gamma_Z m_Z}{s} \text{Im} \left( C_V^{LL} C_Z^{LL\star} - C_V^{LR} C_Z^{RR\star} \right) \Bigg] ~,
\end{multline}

\begin{multline}
I_2^{e^+_R e^-_L}(s) = 8 \Big( |C_\gamma^{LR}|^2 + |C_\gamma^{RL}|^2 \Big) s_W^2 c_W^2 + \frac{8 s_W c_W (1 - 2s_W^2) s^2}{(s-m_Z^2)^2 - \Gamma_Z^2 m_Z^2} \Bigg[ \Big( |C_Z^{LR}|^2 + |C_Z^{RL}|^2 \Big) \frac{(1 - 2 s_W^2)}{4 s_W c_W} \\ + \left( 1 - \frac{m_Z^2}{s} \right) \text{Re} \left( C_\gamma^{LR} C_Z^{LR\star} + C_\gamma^{RL} C_Z^{RL\star} \right) - \frac{\Gamma_Z m_Z}{s} \text{Im} \left( C_\gamma^{LR} C_Z^{LR\star} + C_\gamma^{RL} C_Z^{RL\star} \right) \Bigg] ~.
\end{multline}

\begin{equation}
I^{e^+_R e^-_R}(s) = \bar I^{e^+_L e^-_L}(s) = \frac{s}{m_Z^2} |C_S^{RL}|^2 ~,\qquad 
I^{e^+_L e^-_L}(s) = \bar I^{e^+_R e^-_R}(s) = \frac{s}{m_Z^2} |C_S^{LR}|^2 ~.
\end{equation}

The coefficients $\bar I_0^{e^+_R e^-_L}$, $\bar I_1^{e^+_R e^-_L}$, $\bar I_2^{e^+_R e^-_L}$ and $\bar I_0^{e^+_L e^-_R}$, $\bar I_1^{e^+_L e^-_R}$, $\bar I_2^{e^+_L e^-_R}$ have the same expressions as $ I_0^{e^+_R e^-_L}$, $ I_1^{e^+_R e^-_L}$, $ I_2^{e^+_R e^-_L}$ and $ I_0^{e^+_L e^-_R}$, $ I_1^{e^+_L e^-_R}$, $ I_2^{e^+_L e^-_R}$ respectively, except with an opposite sign for the corresponding imaginary parts in each expression. This reflects the relationship between the different channels under the $CP$ transformation.

As discussed in detail in~\cite{Altmannshofer:2023tsa}, we find that the effects of renormalization group running from the scale of new physics $\Lambda$ to the energy scales at which the linear colliders will run are essentially negligible. Nevertheless, we choose to incorporate the running in a 1-loop leading logarithmic approximation when displaying our results in section~\ref{sec:results}. For the renormalization group running, we use the results from~\cite{Jenkins:2013wua, Alonso:2013hga}.

\section{Expected sensitivities at future linear colliders} \label{sec:sensitivities}

Our analysis of $e^+ e^- \to \tau \mu$ follows the same basic strategy as in~\cite{Dam:2018rfz, Altmannshofer:2023tsa} and focuses on well reconstructable exclusive hadronic tau decays with multiple hadrons in the final state. As detailed in~\cite{Dam:2018rfz, Altmannshofer:2023tsa}, the main source of background is from $e^+e^- \to \tau^+\tau^- \to \tau_{\rm had}\mu \nu \nu$ where one of the produced taus decays leptonically to a muon and neutrinos while the other tau decays hadronically.
For such processes, the momentum of the muon is an extremely powerful discriminator between signal and background. (See~\cite{Jahedi:2024kvi} for alternative analysis strategies which also result in interesting sensitivities.) In the following two subsections we detail how we determine the expected muon momentum distributions of both signal and background events.

\subsection{Muon momentum distribution of the signal} \label{sec:signal}

The momentum of the muon is conveniently parametrized by the variable 
\begin{equation}
    x = \frac{p_\text{det}}{p_\text{beam}} ~,
\end{equation}
where $p_\text{det}$ is the absolute value of the measured muon momentum and $p_\text{beam}$ is the absolute value of the nominal beam momentum $p_\text{beam} = \sqrt{s}/2$, with $\sqrt{s}$ the nominal center-of-mass energy of the collider.
While background events give a broad distribution in the entire range $0 < x \lesssim 1$, we expect the signal to be sharply peaked around $x = 1$ with the width of the peak dictated by the beam energy spread and the momentum resolution of the detector. In addition, initial state radiation (ISR) of photons by the electron and positron beams causes the signal distribution to acquire a tail to lower $x$ values and its peak location to be shifted slightly below a value of~$x=1$.

In Appendix~\ref{appendix}, we derive a closed-form expression for the expected signal distribution in the lab frame, that takes into account all three convolutions with the beam energy spread, ISR effects, and detector resolution in the appropriate order. In practice, we have found it is easier to extract the signal distribution from Monte Carlo simulations
\begin{equation} \label{eq:dsigma_x_sig}
\left(\frac{1}{\sigma}\frac{d\sigma}{dx} \right)_\text{signal} = \left(\frac{1}{\sigma}\frac{d\sigma}{dx} \right)_\text{signal}^\text{MC} ~.
\end{equation}
Our Monte Carlo simulation proceeds in four steps.

\underline{\it Step 1:} We sample two beam momenta from Gaussian distributions with a mean of $\sqrt{s}/2$ and with a width corresponding to the energy spread of the beams. The relevant beam parameters are taken from~\cite{Adolphsen:2013kya, CLICdp:2018cto} and collected in table~\ref{table:info_ILC} for the ILC\footnote{For the ILC, we combine the uncertainties of the energies of the electron and positron beam from~\cite{Adolphsen:2013kya}, to obtain the uncertainty on the center-of-mass energy $$\frac{\delta \sqrt{s}}{\sqrt{s}} = \frac{1}{2} \sqrt{ \left( \frac{\delta E_{e^-}}{E_{e^-}} \right)^2 + \left(\frac{\delta E_{e^+}}{E_{e^+}}\right)^2 }~.$$} and in table~\ref{table:info_CLIC} for CLIC. 

{\setlength{\tabcolsep}{16pt}
\begin{table}[tb] 
\centering
\begin{tabular}{c c c c c c}
\hline\hline
$\sqrt{s}$~[GeV]& $\mathcal L_{\text{int}}$~[ab$^{-1}$] & $\frac{\delta E_{e^-}}{E_{e^-}}~[\%]$ & $\frac{\delta E_{e^+}}{E_{e^+}}~[\%]$ &  $\frac{\delta \sqrt{s}}{\sqrt{s}}~[\%]$ & $\frac{\delta p}{p}$~[\%] \\ [0.5ex]
\hline
250 & 2 & 0.190 & 0.152 & 0.122  & 0.25 \\
350 & 0.2 & 0.158 & 0.100 & 0.093 & 0.35  \\ 
500 & 4 & 0.124 & 0.070 & 0.071 & 0.5  \\
1000 & 8 & 0.083 & 0.043 & 0.047 & 1.0 \\
\hline \hline 
\end{tabular} 
\caption{Center-of-mass energies, integrated luminosities, as well as relevant beam and detector parameters for the ILC. Values are collected from~\cite{Adolphsen:2013kya, ILCInternationalDevelopmentTeam:2022izu}. The parameters for the 1\,TeV run are those corresponding to configuration ``A1''~\cite{Adolphsen:2013kya}.}
\label{table:info_ILC}
\end{table}}
{\setlength{\tabcolsep}{16pt}
\begin{table}[tb] 
\centering
\begin{tabular}{c c c c}
\hline\hline
$\sqrt{s}$~[GeV] & $\mathcal{L}_{\rm int}$~[ab$^{-1}$] & $\frac{\delta \sqrt{s}}{\sqrt{s}}$~[\%] & $\frac{\delta p}{p}$~[\%] \\ [0.5ex]
\hline
380 & 1.5 & 0.35 & 0.38 \\ 
1500 & 2.5 & 0.35& 1.50 \\
3000 & 5.0 & 0.35 & 3.00 \\
\hline \hline 
\end{tabular} 
\caption{Center-of-mass energies, integrated luminosities, as well as relevant beam and detector parameters for CLIC. Values are collected from~\cite{CLICdp:2018cto, Brunner:2022usy}\footnote{Note that the integrated luminosity $\mathcal{L}_{\rm int}=1.5~ \text{ab}^{-1}$ quoted in~\cite{Brunner:2022usy} for the $\sqrt{s}=380$ GeV run is slightly larger than the one of $\mathcal{L}_{\rm int}=1.0~ \text{ab}^{-1}$ quoted in \cite{CLIC:2018fvx,CLICdp:2018cto}.}.}
\label{table:info_CLIC}
\end{table}}

\underline{\it Step 2:} We take into account the ISR effects that we model using the distribution provided in~\cite{NICROSINI1987551}. In particular, the probability distribution function that an electron or positron loses a longitudinal momentum fraction $1-z$ through ISR is given by
\begin{equation}
\label{eq:ISR}
    D(z) = \frac{\beta}{2}(1-z)^{\frac{\beta}{2}-1} \Delta^\prime - \frac{\beta}{4} (1+z) + \frac{\beta^2}{32} \left [(1+z) 
    \Big(3\ln z -4 \ln (1-z) \Big) - \frac{4\ln z}{1-z} -5 -z \right ] ~,
\end{equation}
at $O(\beta^2)$ where 
\begin{equation}
    \beta =  \frac{2\alpha}{\pi} \left [ \ln (s/m_e^2)  - 1 \right ] ~, \qquad \Delta^\prime = 1 + \frac{3\beta}{8} ~.
\end{equation}
After sampling from this distribution, we obtain the electron and positron momenta in the lab frame that we denote with $x_- \sqrt{s}/2$ and $x_+ \sqrt{s} / 2$ respectively, where $x_\pm \neq 1$ due to ISR and beam momentum spread. As a cross-check, we have verified that the above distribution accurately reproduces the behavior of the $e^\pm$ parton distribution functions in $e^\pm$ beams given in~\cite{Garosi:2023bvq} near $z=1$. For a comparison of different approaches~\cite{Kuraev:1985hb, Jadach:2000ir, NICROSINI1987551} in modeling the ISR effects, see~\cite{Greco:2016izi}. Based on the discussion therein, our results are expected to be accurate to the percent level.

\underline{\it Step 3:} We determine the momentum of the muon that is created in the $e^+ e^- \to \tau \mu$ process. In the lab frame, we denote this momentum with $p$ and it is related to the scattering angle $\theta$ in the center-of-mass frame by
\begin{equation}
\label{eq:cos_theta}
    \cos \theta = \frac{1}{(x_- - x_+)} \left( \frac{4p}{\sqrt{s}} - x_- - x_+ \right) ~.
\end{equation}
We thus find the following expression for the differential cross section as a function of the muon momentum in the lab frame
{\everymath{\displaystyle}
\begin{equation}
\label{eq:dsigma_dp_signal}
    \frac{d\sigma}{dp} =  \begin{cases} \frac{4}{\sqrt{s}} \frac{1}{(x_--x_+)} ~ \frac{d\sigma}{d\cos\theta} & \text{for} ~ \text{min}(x_-,x_+) \frac{\sqrt{s}}{2} < p < \text{max}(x_-,x_+) \frac{\sqrt{s}}{2}~, \\
    0 & \text{otherwise}~. \end{cases}
\end{equation}}
The expressions for the angular differential cross-section $d\sigma/d\cos\theta$ is the one given in equation~\eqref{eq:dsigma}. We sample the muon momentum from the differential cross section in~\eqref{eq:dsigma_dp_signal}.

\underline{\it Step 4:} We smear the muon momentum to take into account the momentum resolution of the detector. We smear the momentum using a Gaussian with the detector parameters from~\cite{ILCInternationalDevelopmentTeam:2022izu, CLICdp:2018cto} and collected in table~\ref{table:info_ILC} for the ILC and in table~\ref{table:info_CLIC} for CLIC.\footnote{We determine the ILC and CLIC momentum resolutions quoted in table~\ref{table:info_ILC} and table~\ref{table:info_CLIC} from the inverse momentum resolution stated in both~\cite{ILCInternationalDevelopmentTeam:2022izu} and~\cite{CLICdp:2018cto}, $\delta\left(1/p\right) = 2 \times 10^{-5}$/GeV.}

\begin{figure}[tb]
\centering
\includegraphics[width=0.46\linewidth]{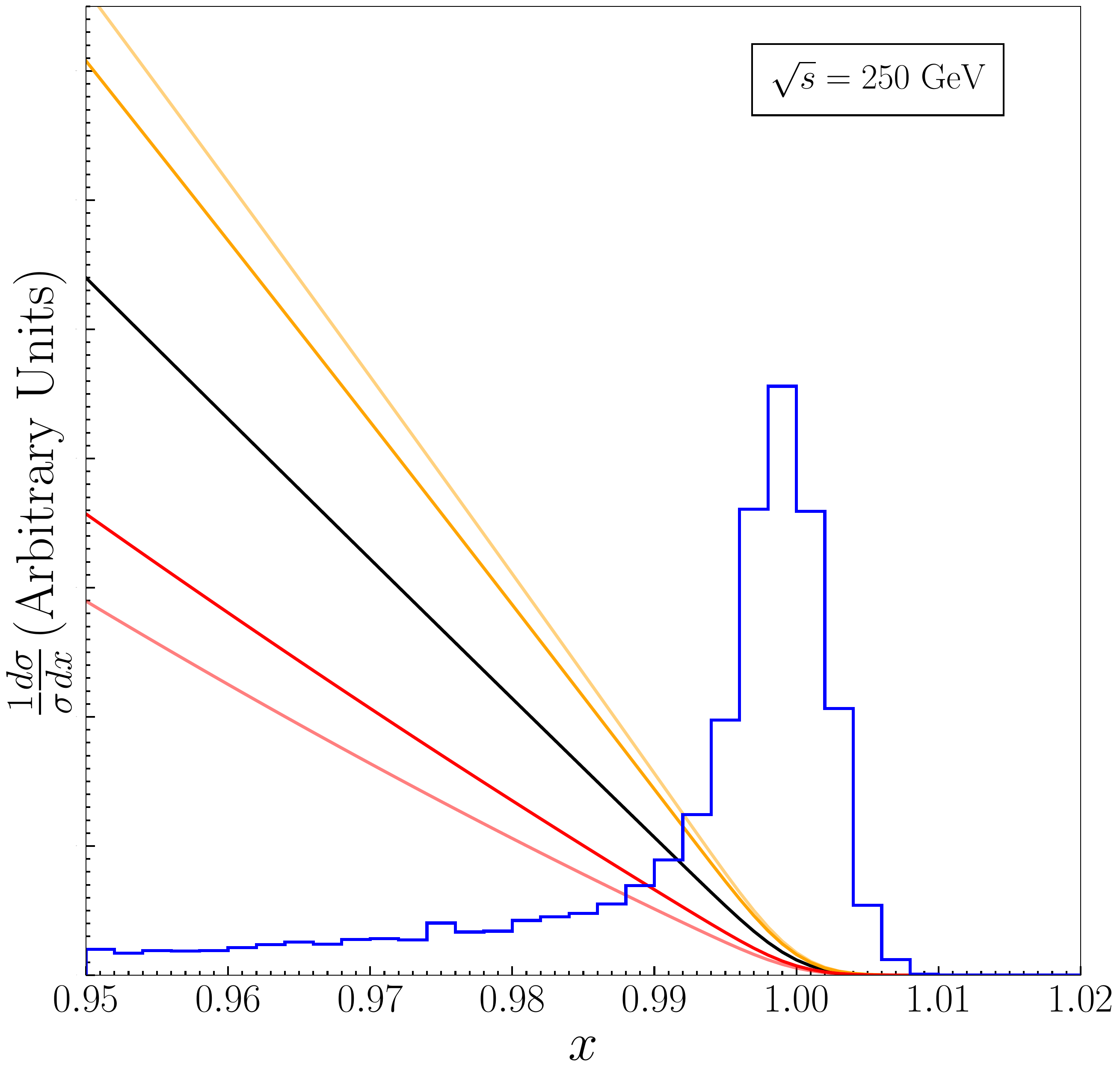} ~~~~
\includegraphics[width=0.46\linewidth]{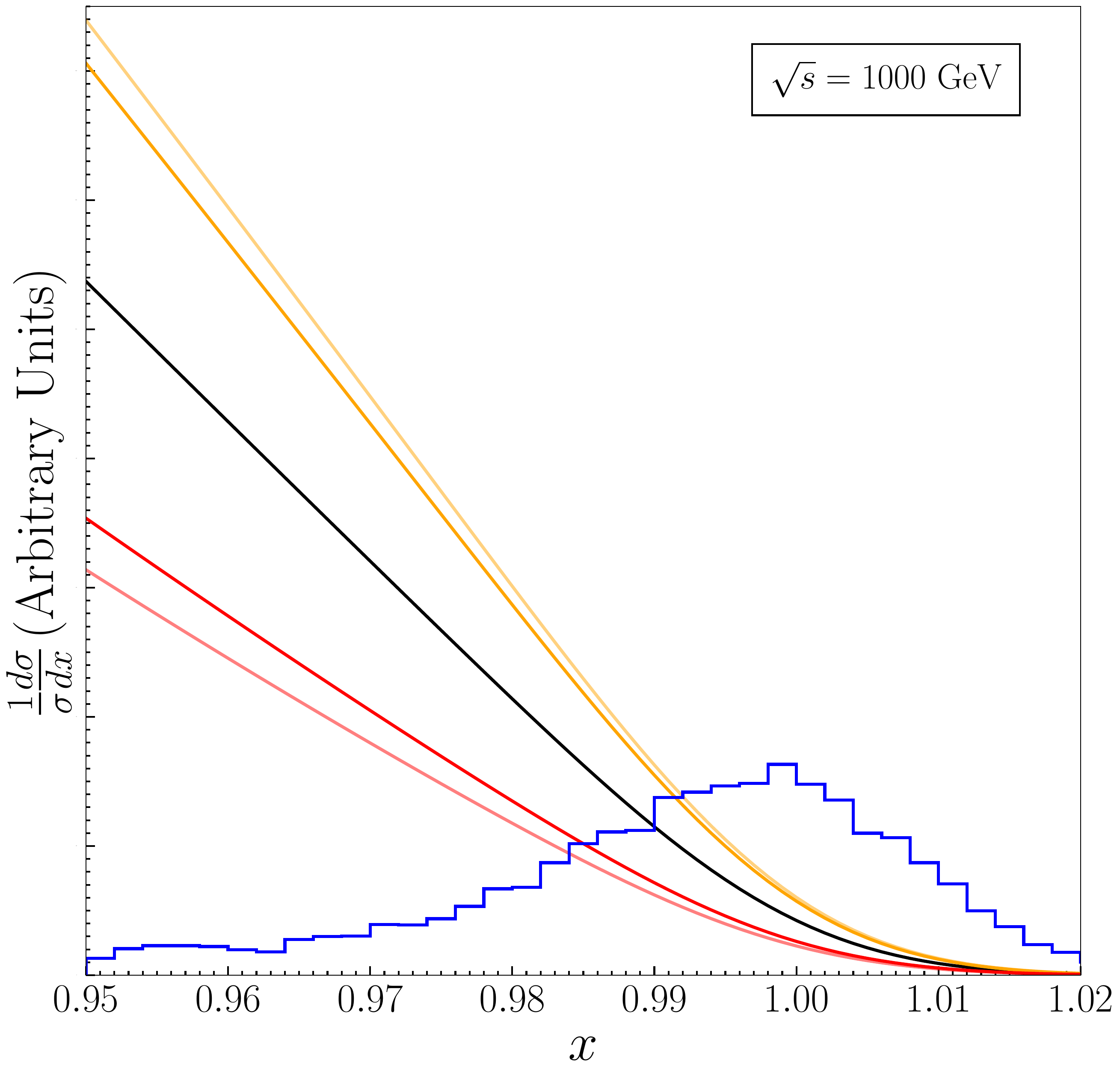}
\caption{Histogram of 10,000 Monte Carlo simulated events for the $e^+ e^- \to \tau \mu$ signal as a function of the muon momentum fraction $x$ when one operator is switched on ({\color{blue} blue}). The left/right plot corresponds to ILC running at a center-of-mass energy of $\sqrt{s} = 250/1000$\,GeV. The background distributions of muons from $e^+e^- \to \tau^+ \tau^- \to \tau \mu \nu \nu$ with beam polarizations $(P_+, P_-) = (0,0)$, $(P_+, P_-) = ({+0.3} / {+0.2},+0.8)$, $(P_+, P_-) = ({-0.3} / {-0.2}, +0.8)$, $(P_+, P_-) = ({-0.3} / {-0.2},-0.8)$ and $(P_+, P_-) = ({+0.3} / {+0.2}, -0.8)$ are shown as solid black, dark red, light red, dark orange and light orange lines respectively. For better visibility, the background distributions have been normalized such that they correspond to $10^3$ times the number of signal events.}
\label{fig:MC}
\end{figure}

\bigskip
We simulate 10,000 events following the procedure described above. Figure \ref{fig:MC} shows in blue two histograms of the muon momentum distribution at the ILC with center-of-mass energies $\sqrt{s} = 250$ GeV (left) and $\sqrt{s} = 1000$ GeV (right) when the Wilson coefficient $C_{V}^{LR}$ is switched on with the remaining Wilson coefficients being zero. We have verified that the shape of the histograms is not affected by which operator we choose to switch on. We have also verified that the signal distribution is to an excellent approximation independent of the beam polarization. 

The width of the signal distribution is mainly determined by the beam energy spread and the detector momentum resolution, which are larger for the 1000 GeV run shown in the right panel than for the 250 GeV run shown in the left panel. Without ISR, the peak of the signal distribution would be centered on $x=1$. The impact of ISR is to create a tail to lower values of $x$ and to slightly shift the peak of the distribution to the left, as evidenced by both plots.

\subsection{Muon momentum distribution of the background}

Various background processes might mimic the $e^+ e^- \to \tau \mu$ signal. To minimize background from $e^+ e^- \to \mu^+ \mu^-$ where a muon is misidentified as a tau lepton, we select only events where the tau decays hadronically to two or three pions.

As discussed in Ref.~\cite{Altmannshofer:2023tsa}, backgrounds due to $e^+ e^- \to W^+ W^- \to \tau_\text{had} \nu \mu \nu$, $e^+ e^- \to W^+ W^- \to \tau_{\rm had} \nu \tau \nu \to \tau_{\rm had} \mu 4\nu$ and $e^+ e^- \to ZZ \to \tau_{\rm had} \tau \nu\nu \to \tau_{\rm had}\mu 4\nu$ are negligible when we apply a cut $x \gtrsim 1$. As long as the intermediate $W$ or $Z$ bosons are on-shell, the momentum of the emitted muon is limited to be strictly below the beam momentum, $x<1$. This conclusion was also verified numerically for off-shell gauge bosons using Monte Carlo simulations, MadGraph5~\cite{Alwall:2014hca} to be specific.\footnote{We found that for the highest center-of-mass energy that could be reached at CLIC, $\sqrt{s} = 3$\,TeV, the $e^+ e^- \to W^+ W^-$ background starts to be non-negligible if one were to impose a cut somewhat below $x = 1$. In our work, we consistently use a cut at $x = 1$ which we found to be sufficient to suppress this background to negligible levels.} 

The dominant remaining background is from the process $e^+ e^- \to \tau_{\rm had} \tau \to \tau_{\rm had} \mu \bar{\nu} \nu $ where one of the taus decays leptonically to a muon and neutrinos, while the other tau decays hadronically to pions. The muon momentum distribution of this decay chain is given by
\begin{equation} \label{eq:sigma_bkg_y}
    \left ( \frac{1}{\sigma}\frac{d\sigma}{dy} \right )_{\rm bkg} = \frac{1}{3} \left [ (5 - 9 y^2 + 4 y^3) + P_\tau (1 - 9 y^2 + 8 y^3)\right ] ~,
\end{equation}
where $y = 2p/\sqrt{s}$, the ratio of the true muon momentum to half the nominal center-of-mass energy.
$P_\tau$ characterizes the asymmetry between right-handed and left-handed taus, which leaves a non-trivial imprint on the muon momentum. For arbitrary beams polarizations, we find that $P_\tau$ is given by
\begin{multline}
\label{eq:Ptau_ee_to_tautau}
 P_\tau = - \Bigg[ \big( 1- P_+ P_- \big) (1-4s_W^2) \Bigg( \frac{1-4 s_W^2 + 8s_W^4}{8 s_W^2 c_W^2} +  1 - \frac{m_Z^2}{s}\Bigg) \\
 + (P_+ - P_-) \Bigg( \frac{(1-4s_W^2)^2}{8 s_W^2 c_W^2} + 1 - \frac{m_Z^2}{s} \Bigg) \Bigg] \\
 \times \Bigg[ \big( 1- P_+ P_- \big) \Bigg( \frac{(1-4 s_W^2 + 8s_W^4)^2}{8 s_W^2 c_W^2} + (1-4s_W^2)^2 \left( 1 - \frac{m_Z^2}{s}\right) + 8s_W^2c_W^2\left( 1 - \frac{m_Z^2}{s}\right)^2 \Bigg) \\
 + (P_+ - P_-) (1-4s_W^2) \Bigg( \frac{1-4 s_W^2 + 8 s_W^4}{8 s_W^2 c_W^2} + 1 - \frac{m_Z^2}{s} \Bigg) \Bigg]^{-1} ~.
\end{multline}

To find the distribution of the background as a function of the measured muon momentum, we follow~\cite{Dam:2018rfz, Altmannshofer:2023tsa} and convolute the distribution in~\eqref{eq:sigma_bkg_y} with a Gaussian of width $(\delta x/x)^2 = (\delta \sqrt{s}/\sqrt{s})^2 + (\delta p/ p)^2$ that takes into account the uncertainties due to the beam energy spread and the detector momentum resolution combined in quadrature
\begin{equation} \label{eq:dsigma_x_bkg}
    \left ( \frac{1}{\sigma}\frac{d\sigma}{dx} \right )_{\rm bkg} \simeq \int dy\, \left(\frac{1}{\sigma} \frac{d\sigma}{dy} \right)_\text{bkg} \frac{1}{\sqrt{2\pi} \delta x} \exp\left(-\frac{(x-y)^2}{2\delta x^2}\right) ~.
\end{equation}
We neglect ISR effects in our modeling of the background. Since ISR radiation tends to shift background events to lower values of $x$, applying a cut on $x$ would be even more effective at suppressing the background if ISR were included. In this sense, the sensitivity estimates presented in the next section are conservative.

We show our analytic estimates of the background distributions in Figure~\ref{fig:MC} and compare them to our Monte Carlo generated signal distributions.
The solid black line shows as reference the unpolarized case $(P_+, P_-) = (0,0)$. The colored lines correspond to the four ILC beam polarizations at 250\,GeV (left plot) and at 1000\,GeV (right plot) as discussed in reference~\cite{ILCInternationalDevelopmentTeam:2022izu}: $(P_+, P_-) = ({-0.3} / {-0.2},-0.8)$ (dark orange), $(P_+, P_-) = ({+0.3} / {+0.2}, -0.8)$ (light orange), $(P_+, P_-) = ({+0.3} / {+0.2},+0.8)$ (dark red), and $(P_+, P_-) = ({-0.3} / {-0.2}, +0.8)$ (light red).
The background distributions are normalized such that their integral over the entire range $0 \lesssim x \lesssim 1$ corresponds to $10^3$ times the integrated signal distribution (shown by the blue histograms). Interestingly, the background shapes depend on the polarization settings. The polarization setting $(P_+, P_-) = ({+0.3} / {+0.2}, -0.8)$ gives the largest tail to high $x$, while $(P_+, P_-) = ({-0.3} / {-0.2}, +0.8)$ gives the smallest tail.

In all cases, only a tiny fraction of the background distributions extend above $x = 1$. Applying a cut $x\gtrsim 1$ eliminates most of the background for all beam polarization settings and is sufficient to almost completely tame this background.

\subsection{Signal and background event numbers}

The total number of expected signal and background events can be written as
\begin{align}
    N_{\rm sig} &= \mathcal{L}_{\rm int} \times\sigma(e^+e^- \to \tau \mu) \times \mathcal{B} (\tau \to \text{had}) \epsilon_{\rm had} \times  \epsilon_{\rm sig}^{x} \times \epsilon_{\rm sig}^{\rm ang} ~, \\
    N_{\rm bkg} &= \mathcal{L}_{\rm int} \times \sigma(e^+e^- \to \tau^+ \tau^-) \times 2 \times \mathcal{B} (\tau \to \text{had}) \epsilon_{\rm had} \times \mathcal{B}(\tau \to \mu \nu\nu) \times \epsilon_{\rm bkg}^{x} \times \epsilon_{\rm bkg}^{\rm ang} ~,
\end{align}
where $\mathcal{L}_{\rm int}$ is the integrated luminosity of the collider, $\mathcal B(\tau \to \mu \nu\nu) \simeq 17.4\%$~\cite{ParticleDataGroup:2024cfk} is the branching ratio of the tau into a muon, and
\begin{multline} \label{eq:tau_had}
\mathcal{B}(\tau\to \text{had})\epsilon_{\rm had} = \mathcal{B}(\tau \to 2\pi \nu)\epsilon_{2\pi} + \mathcal{B}(\tau^\pm \to \pi^\pm 2\pi^0 \nu)\epsilon_{\text{1-prong}} \\ + \mathcal{B}(\tau^\pm \to \pi^\pm \pi^+ \pi^- \nu)\epsilon_{\text{3-prong}} ~,
\end{multline}
refers to the sum of the branching ratios of the tau into 2-pion and 3-pion final states multiplied by their respective detector identification efficiencies. We use the measured tau branching ratios from the PDG~\cite{ParticleDataGroup:2024cfk} and the efficiencies $\epsilon_{2\pi} \simeq 91.6\%, \epsilon_{1-\text{prong}} \simeq 67.5\%$ and $\epsilon_{3-\text{prong}} \simeq 91.1\%$ from the ILC document~\cite{Behnke:2013lya}. We assume that the pion identification efficiencies for the CLIC detectors are the same as those of the ILC in our analysis.

The remaining ingredients to determine the event numbers are the total signal and background cross sections ($\sigma(e^+e^- \to \tau \mu)$ and $\sigma(e^+e^- \to \tau^+ \tau^-)$), the signal and background efficiencies due to the cut on the variable $x$ ($\epsilon_\text{sig}^x$ and $\epsilon_\text{bkg}^x$), and the angular acceptances of the signal and background ($\epsilon_{\rm sig}^{\rm ang}$ and $\epsilon_{\rm bkg}^{\rm ang}$).

\underline{\it Cross sections:} The signal cross section $\sigma(e^+ e^- \to \tau \mu)$ was already given in equation~\eqref{eq:sigma_tot} and for the 
background, we find that the total cross-section for arbitrary beam polarization settings is given by
\begin{multline}
\label{eq:sigma_tot_ee_to_tautau}
 \sigma(e^+ e^- \to \tau^+ \tau^-) = \frac{4 \alpha^2 \pi}{3 s} \Bigg\{ \big( 1- P_+ P_- \big) \\ 
 \times \Bigg[ 1 + \frac{s^2}{(s-m_Z^2)^2 + \Gamma_Z^2 m_Z^2} \Bigg( \frac{(1-4 s_W^2 + 8 s_W^4)^2}{64 s_W^4 c_W^4}
 + \frac{(1- 4 s_W^2)^2}{8 s_W^2 c_W^2} \left( 1 - \frac{m_Z^2}{s}\right) \Bigg) \Bigg] \\
 + \big(P_+ - P_- \big) \frac{s^2}{(s-m_Z^2)^2 + \Gamma_Z^2 m_Z^2}  \left( \frac{1-4 s_W^2 + 8 s_W^4}{8 s_W^2 c_W^2} + 1 - \frac{m_Z^2}{s} \right)\frac{1- 4 s_W^2}{8 s_W^2 c_W^2} \Bigg\} ~.
\end{multline}

\underline{\it Cut efficiencies:} The signal and background efficiencies of our analysis can be found from the $x$ distributions discussed above in equations~\eqref{eq:dsigma_x_sig} and \eqref{eq:dsigma_x_bkg}
\begin{equation}
    \epsilon_{\rm sig}^x = \int_{x_c}^{\infty} dx~ \left(\frac{1}{\sigma}\frac{d\sigma}{dx} \right)_\text{sig}^{\rm MC} ~, \quad \epsilon_{\rm bkg}^x = \int_{x_c}^{\infty} dx~ \left(\frac{1}{\sigma}\frac{d\sigma}{dx} \right)_\text{bkg} ~.
\end{equation}
Throughout our analysis, we consistently apply a cut at $x \geq x_c = 1$, which retains an $O(1)$ fraction of the signal while removing the majority of the background. In principle, the sensitivity to the signal could be further improved by optimizing the cut value, for instance by placing it at the peak of the signal distribution or by determining an optimal cut for each center-of-mass energy and beam polarization configuration individually. Additionally, incorporating other discriminating variables, for example angular distributions or the muon impact parameter, could provide further separation between signal and background. However, such detailed optimizations are beyond the scope of this study, and as a result, our sensitivity estimates should be regarded as conservative.
 
\underline{\it Angular acceptance:} We note that reference~\cite{Behnke:2013lya} states that for the ILC, the muon identification efficiency is above 95\% for the angular range of $10^\circ < \theta < 170^\circ$. We thus estimate the angular acceptances as
\begin{equation}
    \epsilon^{\rm ang}_{\rm bkg/sig} = \frac{1}{\sigma_{\rm bkg/sig}} \int_{10^\circ}^{170^\circ} \frac{d\sigma_\text{bkg/sig}}{d\cos\theta} ~ \sin\theta ~ d\theta ~.
\end{equation}
The angular distribution of the signal can be found in equation~\eqref{eq:dsigma}, while for the background we have 
\begin{equation}
\label{eq:dsigma_ee_to_tautau}
 \frac{d\sigma(e^+ e^- \to \tau^+ \tau^-)}{d\cos\theta} = \frac{3}{8} \sigma(e^+ e^- \to \tau^+ \tau^-) \left( 1 + \cos^2\theta + \frac{8}{3} A_\text{FB} \cos\theta \right) ~.
\end{equation}
with the forward-backward asymmetry for arbitrary beam polarizations
\begin{multline}
\label{eq:AFB_ee_to_tautau}
 A_\text{FB} = \frac{3}{4} \Bigg[ \big( 1- P_+ P_- \big) \Bigg( \frac{(1-4 s_W^2)^2}{8 s_W^2 c_W^2} +  1 - \frac{m_Z^2}{s}\Bigg) \\
 + (P_+ - P_-) (1-4s_W^2) \Bigg( \frac{1-4 s_W^2 + 8 s_W^4}{8 s_W^2 c_W^2} + 1 - \frac{m_Z^2}{s} \Bigg) \Bigg] \\
 \times \Bigg[ \big( 1- P_+ P_- \big) \Bigg( \frac{(1-4 s_W^2 + 8s_W^4)^2}{8 s_W^2 c_W^2} + (1-4s_W^2)^2 \left( 1 - \frac{m_Z^2}{s}\right) + 8s_W^2c_W^2\left( 1 - \frac{m_Z^2}{s}\right)^2 \Bigg) \\
 + (P_+ - P_-) (1-4s_W^2) \Bigg( \frac{1-4 s_W^2 + 8 s_W^4}{8 s_W^2 c_W^2} + 1 - \frac{m_Z^2}{s} \Bigg) \Bigg]^{-1} ~.
\end{multline}
We find that the angular acceptance is at least 98$\%$ for both the signal and background independent of the beam polarizations and of the new physics operator that generates the signal. We adopt the same angular efficiencies for the ILC and CLIC.

With the signal and backround event numbers at hand, we can estimate the new physics sensitivities of $e^+e^- \to \tau\mu$ searches at ILC and CLIC.
We employ the following criterion to estimate the $\sim 2 \sigma$ sensitivity to the new physics signal
\begin{equation} \label{eq:criterion}
    N_{\rm sig} \geq 2 \sqrt{N_{\rm bkg} + N_{\rm sig}} ~. 
\end{equation}

Before discussing our results, a few comments are in order about the assumed center-of-mass energies, the integrated luminosities at each center-of-mass energy, and the respective polarization settings of the ILC and CLIC.\footnote{The recent proposal~\cite{LinearCollider:2025lya} for a linear collider facility at CERN describes similar run parameters. The main differences are an integrated luminosity of $3\,\text{ab}^{-1}$ at $\sqrt{s} = 250$\,GeV and an integrated luminosity of $8\,\text{ab}^{-1}$ at $\sqrt{s} = 550$\,GeV. These higher integrated luminosities would slightly improve our sensitivity estimates.} The ILC run scenarios that we consider are taken from~\cite{ILCInternationalDevelopmentTeam:2022izu} and summarized in table~\ref{table:info_ILC_sens}. At each of the center-of-mass energies of $\sqrt{s} = 250,\,350,\, 500$\,GeV, there are four polarization settings: $(P_+, P_-) = (+0.3,+0.8),\, (P_+, P_-) = (+0.3,-0.8),\, (P_+, P_-) = (-0.3,+0.8),\, (P_+, P_-) = (-0.3, -0.8)$. At the 1\,TeV run the polarization of the positron beam is slightly lower: $(P_+, P_-) = (+0.2,+0.8),\, (P_+, P_-) = (+0.2,-0.8),\, (P_+, P_-) = (-0.2,+0.8),\, (P_+, P_-) = (-0.2, -0.8)$. We denote the four settings with $(++),\, (+-),\, (-+),\, (--)$ for simplicity. For CLIC we use the run parameters from~\cite{Brunner:2022usy,CLICdp:2018cto,CLIC:2018fvx}, summarized in table~\ref{table:info_CLIC_sens}. The positron beam is taken to be unpolarized $P_+ = 0$ for all center-of-mass energies, while the electron beam can be positively polarized $P_- = +0.8$ or negatively polarized
$P_- = -0.8$. We denote the two settings with $(+),\,(-)$.
The third column in tables~\ref{table:info_ILC_sens} and~\ref{table:info_CLIC_sens} shows the fraction of data that is collected in each polarization setting.  

{\setlength{\tabcolsep}{8pt}
\begin{table}[tb] 
\centering
\begin{tabular}{c c c c}
\hline\hline
$\sqrt{s}$~[GeV]& $\mathcal L_\text{int}~[\text{ab}^{-1}]$ & $ \big\{ \mathcal L_\text{int}^{(++)} , \mathcal L_\text{int}^{(+-)} , \mathcal L_\text{int}^{(-+)} , \mathcal L_\text{int}^{(--)} \big\} / \mathcal L_\text{int}$ &$ \big\{ \sigma_\text{sig}^{(++)} , \sigma_\text{sig}^{(+-)} , \sigma_\text{sig}^{(-+)} , \sigma_\text{sig}^{(--)} \big\}~[\text{ab}]$ \\ [0.5ex]
\hline
250 & 2 & $\big\{ 5\%, ~45\%, ~45\%, ~5\% \big\}$ & $\big\{ 580
, ~82, ~70, ~590
\big\}$\\
350 & 0.2 & $ \big\{ 5\%, ~68\%, ~22\%, ~5\% \big\} $ & $\big\{5400
, ~420
, ~1200
, ~5400
\big\}$  \\ 
500 & 4 & $ \big\{ 10\%, ~40\%, ~40\%, ~10\% \big\} $ & $\big\{140
, ~45, ~37, ~140\big\}$  \\
1000 & 8 & $\big\{ 10\%, ~40\%, ~40\%, ~10\% \big\}$ & $\big\{61, ~23, ~18, ~65 \big\}$ \\
\hline \hline 
\end{tabular} 
\caption{ILC run parameters from~\cite{ILCInternationalDevelopmentTeam:2022izu} and sensitivity projections to the $e^+e^- \to \tau \mu$ cross section. The third column shows the fraction of integrated luminosity that will be collected in each of the four polarization settings. The fourth column shows our estimates for the signal cross section sensitivity for each run.}
\label{table:info_ILC_sens}
\end{table}}
{\setlength{\tabcolsep}{8pt}
\begin{table}[tb] 
\centering
\begin{tabular}{c c c c}
\hline\hline
$\sqrt{s}$~[GeV]& $\mathcal L_\text{int}~[\text{ab}^{-1}]$ & $ \big\{ \mathcal L_\text{int}^{(+)} , \mathcal L_\text{int}^{(-)} \big\} / \mathcal L_\text{int}$ &$ \big\{ \sigma_\text{sig}^{(+)} , \sigma_\text{sig}^{(-)} \big\}~[\text{ab}]$ \\ [0.5ex]
\hline
380 & 1.5 & $\big\{ 50\%, ~50\% \big\}$ & $\big\{ 66, ~78\big\}$\\
1500 & 2.5 & $ \big\{ 20\%, ~80\% \big\} $ & $\big\{79, ~27\big\}$  \\ 
3000 & 5.0 & $ \big\{ 20\%, ~80\% \big\} $ & $\big\{39, ~15 \big\}$ \\
\hline \hline 
\end{tabular} 
\caption{CLIC run parameters from~\cite{Brunner:2022usy, CLIC:2018fvx}\footnote{Note that the integrated luminosity $\mathcal{L}_{\rm int}=1.5~ \text{ab}^{-1}$ quoted in~\cite{Brunner:2022usy} for the $\sqrt{s}=380$ GeV run is slightly larger than the one of $\mathcal{L}_{\rm int}=1.0~ \text{ab}^{-1}$ quoted in \cite{CLIC:2018fvx,CLICdp:2018cto}.} and sensitivity projections to the $e^+e^- \to \tau \mu$ cross section. The third column shows the fraction of integrated luminosity that is collected in each of the two polarization settings. The fourth column shows our estimates for the signal cross section sensitivity for each run.}
\label{table:info_CLIC_sens}
\end{table}}

The fourth column in tables~\ref{table:info_ILC_sens} and~\ref{table:info_CLIC_sens} shows the sensitivity to the $e^+ e^- \to \tau \mu$ cross section that we find for the various runs of ILC and CLIC using equation~\eqref{eq:criterion}. Generically, the smallest cross sections can be probed at the highest energies, and we find sensitivities at the level of few 10's of atto-barns. 

\section{Results and Discussion} \label{sec:results}

We now translate the sensitivities to $e^+ e^- \to \tau \mu$ cross sections into sensitivities to new physics Wilson coefficients of the SMEFT. To illustrate the complementarity of runs with different polarization settings we discuss two benchmark cases where we switch on two coefficients at a time:
\begin{itemize}
\item[(a)] the SMEFT coefficients $(C_{le})_{\mu e e \tau}$ and $(C_{ee})_{ee\mu\tau}$, which correspond to a scalar current and a right-handed vector current for the electrons, respectively;
\item[(b)] the SMEFT coefficients $(C_{le})_{e e \mu \tau}$ and $(C_{le})_{\mu\tau e e}$, which correspond to a left-handed vector current and right-handed vector current for the electrons, respectively.
\end{itemize}

\begin{figure}[tb]
\centering
\includegraphics[width=0.46\linewidth]{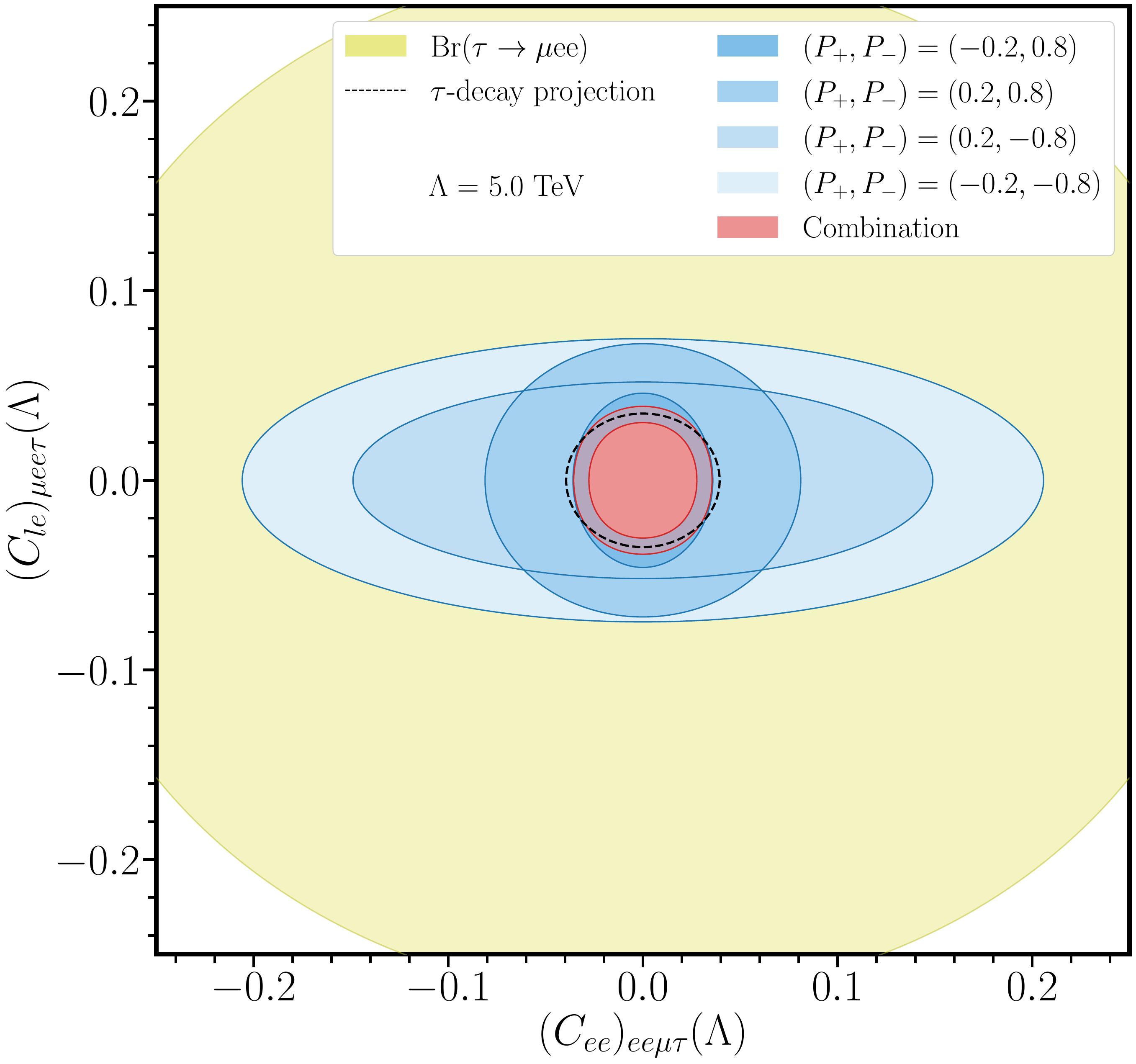} ~~~~
\includegraphics[width=0.46\linewidth]{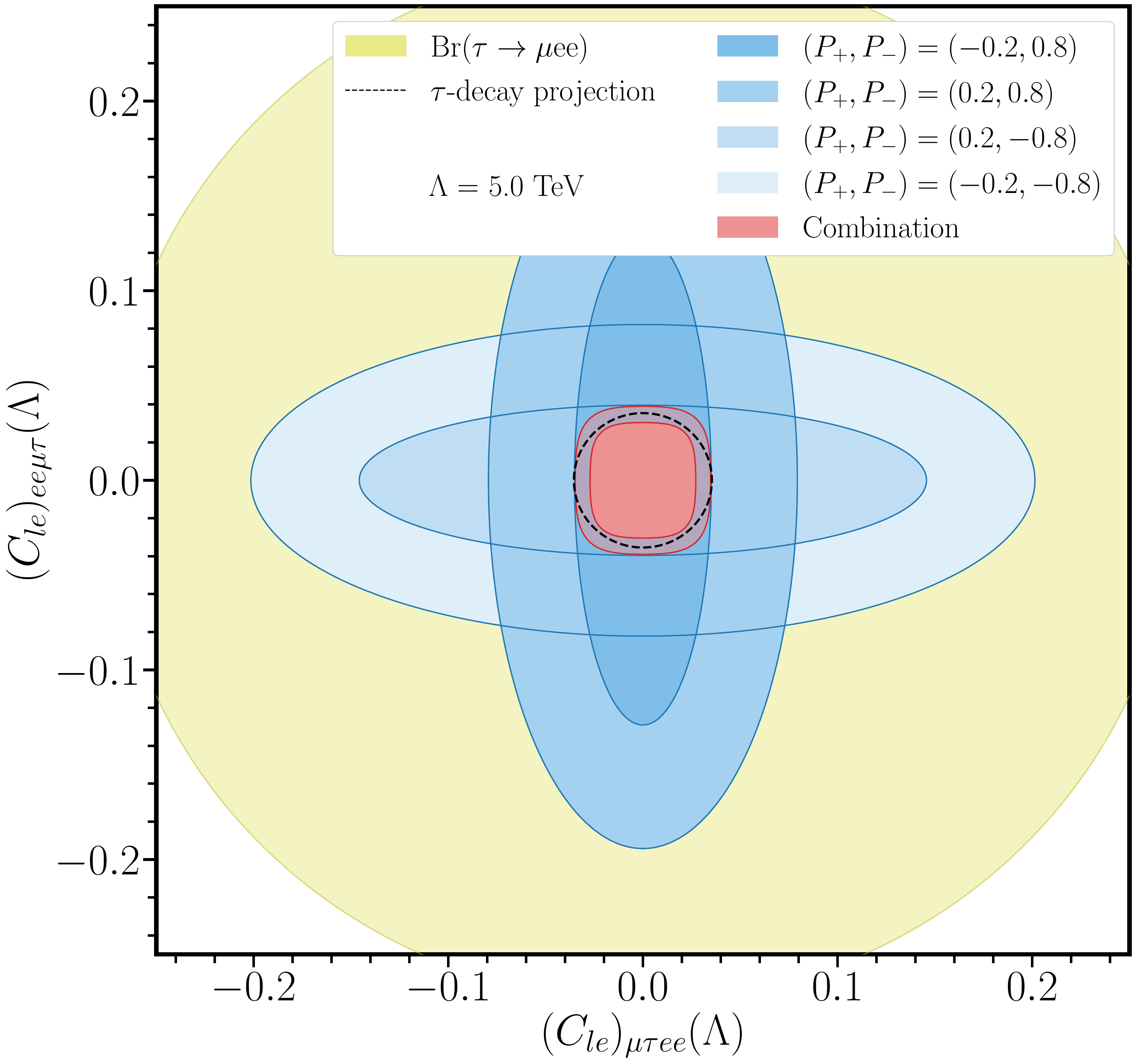}
\caption{Sensitivity to pairs of SMEFT coefficients at the $\sqrt{s} = 1$\,TeV run at ILC. In different shades of blue are our $2\sigma$ sensitivity projections for the different polarizations of the $e^+/e^-$ beams in the process $e^+e^-\to\tau\mu$. The lighter red and darker red regions are the  $2\sigma$ and $1\sigma$ combinations of these constraints. The yellow region shows the current $2\sigma$ constraints from searches for the $\tau\to\mu e^+ e^-$ decay at BaBar and Belle~\cite{Hayasaka:2010np, BaBar:2010axs}. The solid dashed line is the corresponding expected sensitivity projection of Belle II~\cite{Belle-II:2018jsg}. The UV scale is taken to be $\Lambda = 5.0$ TeV.}
\label{fig:Wilson_complementary_ILC}
\end{figure}
\begin{figure}[tb]
\centering
\includegraphics[width=0.46\linewidth]{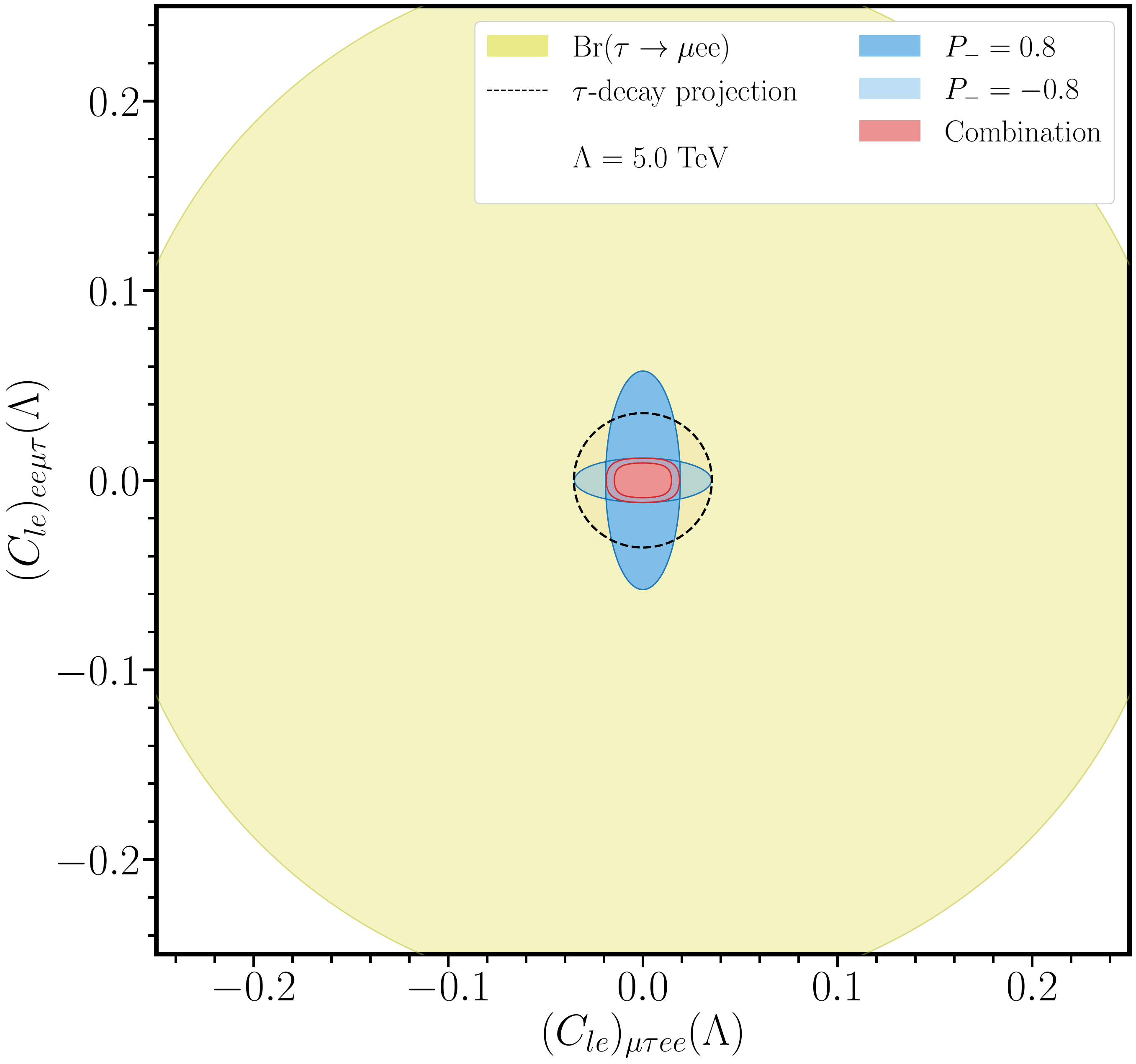} ~~~~
\includegraphics[width=0.46\linewidth]{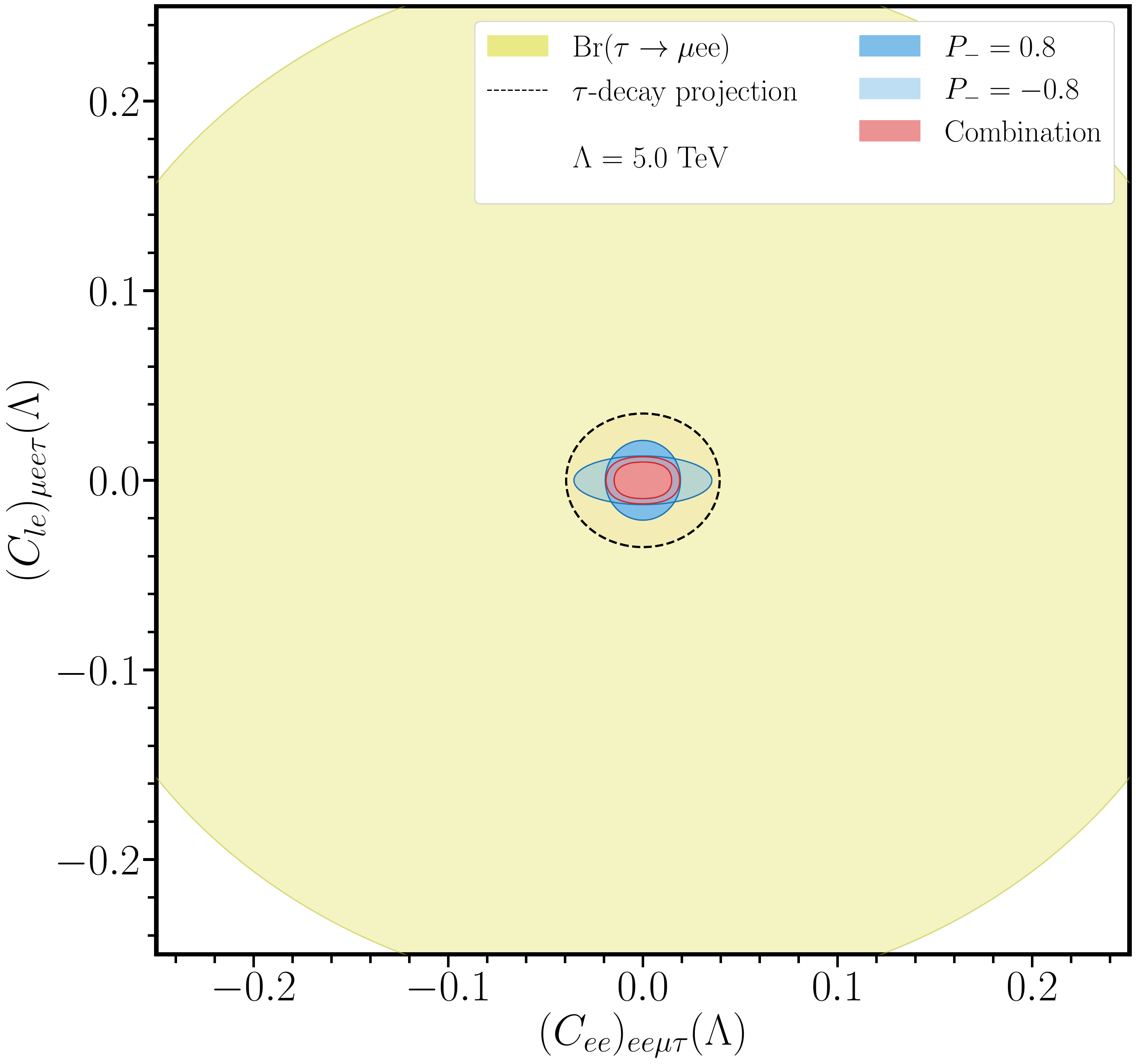}
\caption{Sensitivity to pairs of SMEFT coefficients at the $\sqrt{s} = 3$\,TeV run at CLIC. In different shades of blue are our $2\sigma$ sensitivity projections for the different polarizations of the $e^+/e^-$ beams in the process $e^+e^-\to\tau\mu$. The lighter red and darker red regions are the  $2\sigma$ and $1\sigma$ combinations of these constraints. The yellow region shows the current $2\sigma$ constraints from searches for the $\tau\to\mu e^+ e^-$ decay at BaBar and Belle~\cite{Hayasaka:2010np, BaBar:2010axs}. The solid dashed line is the corresponding expected sensitivity projection of Belle II~\cite{Belle-II:2018jsg}. The UV scale is taken to be $\Lambda = 5.0$ TeV.} 
\label{fig:Wilson_complementary_CLIC}
\end{figure}

Figure~\ref{fig:Wilson_complementary_ILC} shows the sensitivity of the $\sqrt{s}= 1$\,TeV run of the ILC to the Wilson coefficients of case (a) in the plot on the left and those of case (b) in the plot on the right. The corresponding plots for the $\sqrt{s} = 3$\,TeV at CLIC are shown in Figure~\ref{fig:Wilson_complementary_CLIC}.

The $2\sigma$ sensitivities for the runs at the different polarization settings are shown with the shaded blue regions. As discussed in the previous section, the ILC at $\sqrt{s} = 1$\,TeV is proposed to operate with the four polarization configurations $(P_+, P_-) = (-0.2,+0.8)$, $(P_+, P_-) = (+0.2,+0.8)$, $(P_+, P_-) = (+0.2,-0.8)$ and $(P_+, P_-) = (-0.2,-0.8)$, whereas CLIC only has two different setting for the $e^-$ beam i.e. $P_- = +0.8$ and $P_- = -0.8$ with the positron beam remaining unpolarized. 
The $1\sigma$ and $2\sigma$ combinations of the runs with different polarizations are shown by the dark and light red regions. Different polarizations of the electron and positron beams show clear complementarity in probing operators with different chirality structure of the leptons. 

To emphasize the impact of the higher center-of-mass energy that can be achieved at CLIC, we show the plots in figures~\ref{fig:Wilson_complementary_ILC} and~\ref{fig:Wilson_complementary_CLIC} with the same axes scales. The sensitivity to the four-fermion contact interactions is significantly improved at CLIC, as expected from the scaling of the new physics cross section with $s$. 

For comparison, all plots also show the current constraints on the Wilson coefficients from the limits on the flavor changing tau decays $\tau \to \mu e^+ e^-$ from the $B$-factories~\cite{Hayasaka:2010np, BaBar:2010axs}. The yellow region corresponds to the $2\sigma$ bound that is based on the Belle result\footnote{To translate to a $2\sigma$ bound, we assume the distribution is a Gaussian with a mean of zero.}~\cite{Hayasaka:2010np}
\begin{equation}
\text{BR}(\tau \to \mu e^+ e^-) < 1.8\times 10^{-8} \quad (\text{at} ~ 90\% ~ \text{C.L.} )~.
\end{equation}
Belle~II is expected to improve the sensitivity to lepton flavor-violating tau decays by more than an order of magnitude~\cite{Belle-II:2018jsg}. The corresponding Belle~II sensitivity is shown by the dashed black lines. Future circular $e^+ e^-$ machines running on the $Z$-pole are anticipated to improve the sensitivity to $\tau \to \mu e^+ e^-$ even further~\cite{Bernardi:2022hny, Ai:2024nmn, FCC:2025lpp}. However, we are not aware of quantitative sensitivity projections for the $\tau \to \mu e^+e^-$ decay, and therefore we only show the Belle~II projection. 
The expected sensitivity of Belle~II is comparable to the one of the $1$\,TeV run at the ILC, but cannot compete with a $3$\,TeV run at CLIC. 

\begin{figure}[tb]
\centering
\includegraphics[width = \linewidth]{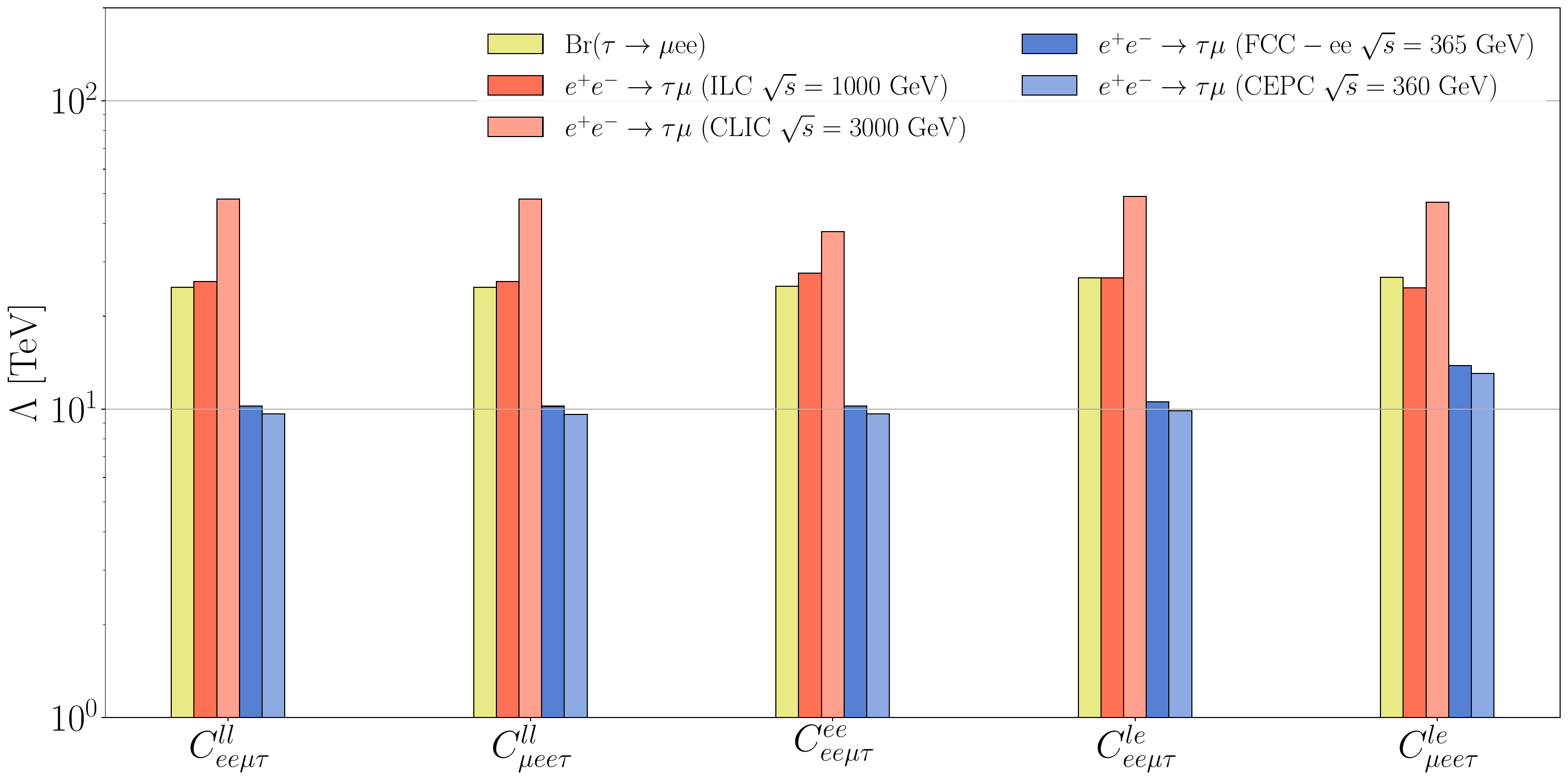}
\caption{Sensitivity to the new physics scale $\Lambda$ from LFV tau decays $\tau\to\mu e^+ e^-$, and searches for $\tau \mu$ production in $e^+e^-$ collisions. Each SMEFT Wilson coefficient is turned on individually at the scale $\Lambda$, i.e. $C_i(\Lambda)=1$, with all others set to zero. Only future projections are displayed. The red bars (ILC and CLIC) are based on the sensitivities we derived in this work, while the blue bars (FCC-ee and CEPC) are updates of the sensitivities we found in Ref.~\cite{Altmannshofer:2023tsa}.} 
\label{fig:bar_chart}
\end{figure}

Figure \ref{fig:bar_chart} shows the sensitivity projections to the new physics scale $\Lambda$ if only one four-fermion SMEFT coefficient is switched on at a time to a value of unity, with all other SMEFT coefficients being set to zero. The yellow bar is the expected sensitivity from searches for $\tau \to \mu e^+ e^-$ at Belle~II~\cite{Belle-II:2018jsg}, the red bars are the expected sensitivities of the ILC running $\sqrt{s} = 1$\,TeV and of CLIC running at $\sqrt{s} = 3$\,TeV. To obtain the shown ILC and CLIC sensitivities, we have combined all proposed runs with different polarization settings. For comparison, we also show with the blue bars the projected sensitivities of $e^+e^- \to \tau \mu$ searches at the FCC-ee and the CEPC from~\cite{Altmannshofer:2023tsa} that we have updated in two ways: (i) we have taken into account ISR effects of the signal analogous to our discussion in section~\ref{sec:signal}; (ii) instead of including hadronic tau decays with two, three, and four pions and a uniform detection efficiency of $25\%$ as in Ref.~\cite{Altmannshofer:2023tsa}, we use the hadronic branching ratios and identification efficiency as in equation~\eqref{eq:tau_had} assuming the one at FCC-ee and CEPC is approximately the same as the one at ILC.

We confirm our estimates from~\cite{Altmannshofer:2023tsa} and find that FCC-ee and CEPC have sensitivities to new physics scales of four-fermion operators of around $10$\,TeV, slightly worse than rare tau decay searches at Belle~II. Searches for $e^+e^- \to \tau \mu$ at the ILC can reach scales up to $\sim 25$\,TeV. The best sensitivity can be achieved at CLIC, around $\Lambda \sim 50$\,TeV. 

\section{Conclusions} \label{sec:conclusions}

In this paper, we have investigated the sensitivity of future linear electron-positron colliders, specifically the ILC and CLIC, for new lepton flavor-violating physics in the $\tau\mu$ sector, focusing on the non-resonant $e^+ e^- \to \tau \mu$ process. We have highlighted the significance of the muon momentum as a powerful discriminating observable to efficiently suppress Standard Model backgrounds, enabling a clean and robust search for the LFV signal in high-energy collisions.

Our analysis is performed within the framework of the Standard Model Effective Field Theory and shows that the sensitivity projections of ILC and CLIC to four-fermion contact interactions are at least competitive with, and in some cases surpass, the expected sensitivities from low-energy $\tau$ decay searches at Belle~II. This is due to the linear growth of the $e^+ e^- \to \tau \mu$ cross section with the squared center-of-mass energy $s$. In particular, at a center-of-mass energy of $\sqrt{s} = 3$ TeV, CLIC could provide the most stringent constraints on the relevant SMEFT operators, significantly extending the reach for LFV interactions in the $\tau\mu$ sector to new physics scales of up to $\sim 50$\,TeV.
An important feature of linear colliders that we have emphasized is their ability to employ polarized electron and positron beams. This capability allows us to disentangle SMEFT operators with different chirality structures.

Our study underscores the complementarity between low-energy precision experiments and high-energy collider searches in the ongoing exploration of lepton flavor violation. While rare decay searches continue to set important constraints, high-energy processes like $e^+e^- \to \tau\mu$ offer a distinct and, in some cases, superior discovery opportunity, especially for scenarios where new physics contributions grow with energy. This highlights the importance of future linear colliders such as the ILC and CLIC as useful tools in the broader program of testing the flavor structure of the Standard Model and searching for signals of new physics.

\section*{Acknowledgements} 

We thank Zoltan Ligeti for useful comments.
The research of W.A. and P.M. is supported by the U.S. Department of Energy grant number DE-SC0010107.

\begin{appendix}
\section{Analytic Expression for the Signal Muon Momentum Distribution} \label{appendix}

In this Appendix, we give a closed-form expression for the differential $e^+ e^- \to \tau \mu$ cross-section as a function of the muon momentum. We take into account the beam energy spread, initial state radiation (ISR) and finite detector resolution effects.

We need to consider the following relevant momenta 
$$ p_\text{beam} ~,\quad p_\text{beam}^+ ~,\quad p_\text{beam}^- ~,\quad p_{\rm ISR}^+ ~,\quad p_{\rm ISR}^- ~,\quad p ~,\quad p_\text{det} ~,$$ 
where $p_\text{beam} = \sqrt{s}/2$ is the nominal beam momentum, $p_\text{beam}^\pm$ are the electron and positron beam momenta taking into account the beam energy spread, $p_{\rm ISR}^\pm$ are the momenta of the electrons and positrons after ISR, $p$ is the momentum of the muon produced from the collision, and $p_\text{det}$ is the muon momentum observed by the detector. 

It will be convenient to introduce the following ratios of momenta
\begin{equation}
u_\pm = \frac{p_\text{beam}^\pm}{p_\text{beam}} = \frac{2 p_\text{beam}^\pm}{\sqrt{s}} ~,\quad x_\pm = \frac{p_\text{ISR}^\pm}{p_\text{beam}}= \frac{2 p_\text{ISR}^\pm}{\sqrt{s}} ~, \quad \omega = \frac{p_\text{det}}{p_\mu} ~,\quad x = \frac{p_\text{det}}{p_\text{beam}} = \frac{2 p_\text{det}}{\sqrt{s}} ~.
\end{equation}

We find that the differential cross section as a function of $x$ can be expressed in the following way
\begin{multline}
\label{eq:convolution}
    \frac{1}{\sigma}\frac{d\sigma}{d x} = \int \frac{du_+}{u_+} \int \frac{du_-}{u_-}  \int d\omega \int dx_+ \int dx_- ~ f_\text{beam}(u_+) f_\text{beam}(u_-) f_\text{det}(\omega) \\ D\left(\frac{x_+}{u_+}\right) D\left(\frac{x_-}{u_-}\right) \frac{2}{x_- - x_+} \frac{1}{\sigma} \frac{d\sigma}{d\cos\theta}(\hat s, \cos\theta) ~,
\end{multline}
where $\hat s = x_+ x_- s$, $\cos\theta = (2x/\omega - x_- - x_+)/(x_- - x_+)$, $d\sigma/d\cos\theta$ is given in~\eqref{eq:dsigma}, and we have the constraint $\text{min}(x_-,x_+) < x/\omega < \text{max}(x_-,x_+)$.  

In the above convolution, $f_\text{beam}$ and $f_\text{det}$ are Gaussian distributions that account for the beam energy spread and finite detector resolution effects respectively, and are given by
\begin{equation}
    f_\text{beam} (u_\pm) = \frac{1}{\sqrt{2\pi} \delta u} \exp\left[-\frac{(1- u_\pm)^2}{2\delta u^2} \right] ~,\quad
    f_\text{det} (\omega) = \frac{1}{\sqrt{2\pi} \delta\omega} \exp\left[-\frac{(1- \omega)^2}{2\delta \omega^2} \right] ~,
\end{equation}
with the beam spread $\delta u = \delta p_\text{beam}^\pm/p_\text{beam}$ and the detector resolution $\delta \omega = \delta p_\text{det} / p$. 

The convolutions with the distribution functions $D$ given in~\eqref{eq:ISR} account for the ISR by the electron and positron. As all the distribution functions are sharply peaked, we find it in practice challenging to perform the convolutions numerically. Therefore, we have determined the differential cross section $\frac{1}{\sigma}\frac{d\sigma}{d x}$ using a Monte Carlo procedure, as described in section~\ref{sec:signal}.  

\end{appendix}

\bibliography{bibliography}

\begin{thebibliography}{65}%
\makeatletter
\providecommand \@ifxundefined [1]{%
 \@ifx{#1\undefined}
}%
\providecommand \@ifnum [1]{%
 \ifnum #1\expandafter \@firstoftwo
 \else \expandafter \@secondoftwo
 \fi
}%
\providecommand \@ifx [1]{%
 \ifx #1\expandafter \@firstoftwo
 \else \expandafter \@secondoftwo
 \fi
}%
\providecommand \natexlab [1]{#1}%
\providecommand \enquote  [1]{``#1''}%
\providecommand \bibnamefont  [1]{#1}%
\providecommand \bibfnamefont [1]{#1}%
\providecommand \citenamefont [1]{#1}%
\providecommand \href@noop [0]{\@secondoftwo}%
\providecommand \href [0]{\begingroup \@sanitize@url \@href}%
\providecommand \@href[1]{\@@startlink{#1}\@@href}%
\providecommand \@@href[1]{\endgroup#1\@@endlink}%
\providecommand \@sanitize@url [0]{\catcode `\\12\catcode `\$12\catcode
  `\&12\catcode `\#12\catcode `\^12\catcode `\_12\catcode `\%12\relax}%
\providecommand \@@startlink[1]{}%
\providecommand \@@endlink[0]{}%
\providecommand \url  [0]{\begingroup\@sanitize@url \@url }%
\providecommand \@url [1]{\endgroup\@href {#1}{\urlprefix }}%
\providecommand \urlprefix  [0]{URL }%
\providecommand \Eprint [0]{\href }%
\providecommand \doibase [0]{http://dx.doi.org/}%
\providecommand \selectlanguage [0]{\@gobble}%
\providecommand \bibinfo  [0]{\@secondoftwo}%
\providecommand \bibfield  [0]{\@secondoftwo}%
\providecommand \translation [1]{[#1]}%
\providecommand \BibitemOpen [0]{}%
\providecommand \bibitemStop [0]{}%
\providecommand \bibitemNoStop [0]{.\EOS\space}%
\providecommand \EOS [0]{\spacefactor3000\relax}%
\providecommand \BibitemShut  [1]{\csname bibitem#1\endcsname}%
\let\auto@bib@innerbib\@empty
\bibitem [{\citenamefont {Marciano}\ and\ \citenamefont
  {Sanda}(1977)}]{Marciano:1977wx}%
  \BibitemOpen
  \bibfield  {author} {\bibinfo {author} {\bibfnamefont {W.~J.}\ \bibnamefont
  {Marciano}}\ and\ \bibinfo {author} {\bibfnamefont {A.~I.}\ \bibnamefont
  {Sanda}},\ }\bibfield  {title} {\enquote {\bibinfo {title} {{Exotic Decays of
  the Muon and Heavy Leptons in Gauge Theories}},}\ }\href {\doibase
  10.1016/0370-2693(77)90377-X} {\bibfield  {journal} {\bibinfo  {journal}
  {Phys. Lett. B}\ }\textbf {\bibinfo {volume} {67}},\ \bibinfo {pages}
  {303--305} (\bibinfo {year} {1977})}\BibitemShut {NoStop}%
\bibitem [{\citenamefont {Petcov}(1977)}]{Petcov:1976ff}%
  \BibitemOpen
  \bibfield  {author} {\bibinfo {author} {\bibfnamefont {S.~T.}\ \bibnamefont
  {Petcov}},\ }\bibfield  {title} {\enquote {\bibinfo {title} {{The Processes
  $\mu \rightarrow e + \gamma, \mu \rightarrow e + \overline{e}, \nu'
  \rightarrow \nu + \gamma$ in the Weinberg-Salam Model with Neutrino
  Mixing}},}\ }\href@noop {} {\bibfield  {journal} {\bibinfo  {journal} {Sov.
  J. Nucl. Phys.}\ }\textbf {\bibinfo {volume} {25}},\ \bibinfo {pages} {340}
  (\bibinfo {year} {1977})},\ \bibinfo {note} {[Erratum: Sov.J.Nucl.Phys. 25,
  698 (1977), Erratum: Yad.Fiz. 25, 1336 (1977)]}\BibitemShut {NoStop}%
\bibitem [{\citenamefont {Lee}\ and\ \citenamefont
  {Shrock}(1977)}]{Lee:1977tib}%
  \BibitemOpen
  \bibfield  {author} {\bibinfo {author} {\bibfnamefont {Benjamin~W.}\
  \bibnamefont {Lee}}\ and\ \bibinfo {author} {\bibfnamefont {Robert~E.}\
  \bibnamefont {Shrock}},\ }\bibfield  {title} {\enquote {\bibinfo {title}
  {{Natural Suppression of Symmetry Violation in Gauge Theories: Muon - Lepton
  and Electron Lepton Number Nonconservation}},}\ }\href {\doibase
  10.1103/PhysRevD.16.1444} {\bibfield  {journal} {\bibinfo  {journal} {Phys.
  Rev. D}\ }\textbf {\bibinfo {volume} {16}},\ \bibinfo {pages} {1444}
  (\bibinfo {year} {1977})}\BibitemShut {NoStop}%
\bibitem [{\citenamefont {Calibbi}\ and\ \citenamefont
  {Signorelli}(2018)}]{Calibbi:2017uvl}%
  \BibitemOpen
  \bibfield  {author} {\bibinfo {author} {\bibfnamefont {Lorenzo}\ \bibnamefont
  {Calibbi}}\ and\ \bibinfo {author} {\bibfnamefont {Giovanni}\ \bibnamefont
  {Signorelli}},\ }\bibfield  {title} {\enquote {\bibinfo {title} {{Charged
  Lepton Flavour Violation: An Experimental and Theoretical Introduction}},}\
  }\href {\doibase 10.1393/ncr/i2018-10144-0} {\bibfield  {journal} {\bibinfo
  {journal} {Riv. Nuovo Cim.}\ }\textbf {\bibinfo {volume} {41}},\ \bibinfo
  {pages} {71--174} (\bibinfo {year} {2018})},\ \Eprint
  {http://arxiv.org/abs/1709.00294} {arXiv:1709.00294 [hep-ph]} \BibitemShut
  {NoStop}%
\bibitem [{\citenamefont {Baldini}\ \emph {et~al.}(2018)\citenamefont {Baldini}
  \emph {et~al.}}]{MEGII:2018kmf}%
  \BibitemOpen
  \bibfield  {author} {\bibinfo {author} {\bibfnamefont {A.~M.}\ \bibnamefont
  {Baldini}} \emph {et~al.} (\bibinfo {collaboration} {MEG II}),\ }\bibfield
  {title} {\enquote {\bibinfo {title} {{The design of the MEG II
  experiment}},}\ }\href {\doibase 10.1140/epjc/s10052-018-5845-6} {\bibfield
  {journal} {\bibinfo  {journal} {Eur. Phys. J. C}\ }\textbf {\bibinfo {volume}
  {78}},\ \bibinfo {pages} {380} (\bibinfo {year} {2018})},\ \Eprint
  {http://arxiv.org/abs/1801.04688} {arXiv:1801.04688 [physics.ins-det]}
  \BibitemShut {NoStop}%
\bibitem [{\citenamefont {Abramishvili}\ \emph {et~al.}(2020)\citenamefont
  {Abramishvili} \emph {et~al.}}]{COMET:2018auw}%
  \BibitemOpen
  \bibfield  {author} {\bibinfo {author} {\bibfnamefont {R.}~\bibnamefont
  {Abramishvili}} \emph {et~al.} (\bibinfo {collaboration} {COMET}),\
  }\bibfield  {title} {\enquote {\bibinfo {title} {{COMET Phase-I Technical
  Design Report}},}\ }\href {\doibase 10.1093/ptep/ptz125} {\bibfield
  {journal} {\bibinfo  {journal} {PTEP}\ }\textbf {\bibinfo {volume} {2020}},\
  \bibinfo {pages} {033C01} (\bibinfo {year} {2020})},\ \Eprint
  {http://arxiv.org/abs/1812.09018} {arXiv:1812.09018 [physics.ins-det]}
  \BibitemShut {NoStop}%
\bibitem [{\citenamefont {Bernstein}(2019)}]{Bernstein:2019fyh}%
  \BibitemOpen
  \bibfield  {author} {\bibinfo {author} {\bibfnamefont {R.~H.}\ \bibnamefont
  {Bernstein}} (\bibinfo {collaboration} {Mu2e}),\ }\bibfield  {title}
  {\enquote {\bibinfo {title} {{The Mu2e Experiment}},}\ }\href {\doibase
  10.3389/fphy.2019.00001} {\bibfield  {journal} {\bibinfo  {journal} {Front.
  in Phys.}\ }\textbf {\bibinfo {volume} {7}},\ \bibinfo {pages} {1} (\bibinfo
  {year} {2019})},\ \Eprint {http://arxiv.org/abs/1901.11099} {arXiv:1901.11099
  [physics.ins-det]} \BibitemShut {NoStop}%
\bibitem [{\citenamefont {Arndt}\ \emph {et~al.}(2021)\citenamefont {Arndt}
  \emph {et~al.}}]{Mu3e:2020gyw}%
  \BibitemOpen
  \bibfield  {author} {\bibinfo {author} {\bibfnamefont {K.}~\bibnamefont
  {Arndt}} \emph {et~al.} (\bibinfo {collaboration} {Mu3e}),\ }\bibfield
  {title} {\enquote {\bibinfo {title} {{Technical design of the phase I Mu3e
  experiment}},}\ }\href {\doibase 10.1016/j.nima.2021.165679} {\bibfield
  {journal} {\bibinfo  {journal} {Nucl. Instrum. Meth. A}\ }\textbf {\bibinfo
  {volume} {1014}},\ \bibinfo {pages} {165679} (\bibinfo {year} {2021})},\
  \Eprint {http://arxiv.org/abs/2009.11690} {arXiv:2009.11690
  [physics.ins-det]} \BibitemShut {NoStop}%
\bibitem [{\citenamefont {Afanaciev}\ \emph {et~al.}(2025)\citenamefont
  {Afanaciev} \emph {et~al.}}]{MEGII:2025gzr}%
  \BibitemOpen
  \bibfield  {author} {\bibinfo {author} {\bibfnamefont {K.}~\bibnamefont
  {Afanaciev}} \emph {et~al.} (\bibinfo {collaboration} {MEG II}),\ }\bibfield
  {title} {\enquote {\bibinfo {title} {{New limit on the $\mu^\pm \to e^\pm
  \gamma$ decay with the MEG II experiment}},}\ }\href@noop {} {\  (\bibinfo
  {year} {2025})},\ \Eprint {http://arxiv.org/abs/2504.15711} {arXiv:2504.15711
  [hep-ex]} \BibitemShut {NoStop}%
\bibitem [{\citenamefont {Altmannshofer}\ \emph {et~al.}(2019)\citenamefont
  {Altmannshofer} \emph {et~al.}}]{Belle-II:2018jsg}%
  \BibitemOpen
  \bibfield  {author} {\bibinfo {author} {\bibfnamefont {W.}~\bibnamefont
  {Altmannshofer}} \emph {et~al.} (\bibinfo {collaboration} {Belle-II}),\
  }\bibfield  {title} {\enquote {\bibinfo {title} {{The Belle II Physics
  Book}},}\ }\href {\doibase 10.1093/ptep/ptz106} {\bibfield  {journal}
  {\bibinfo  {journal} {PTEP}\ }\textbf {\bibinfo {volume} {2019}},\ \bibinfo
  {pages} {123C01} (\bibinfo {year} {2019})},\ \bibinfo {note} {[Erratum: PTEP
  2020, 029201 (2020)]},\ \Eprint {http://arxiv.org/abs/1808.10567}
  {arXiv:1808.10567 [hep-ex]} \BibitemShut {NoStop}%
\bibitem [{\citenamefont {Delepine}\ and\ \citenamefont
  {Vissani}(2001)}]{Delepine:2001di}%
  \BibitemOpen
  \bibfield  {author} {\bibinfo {author} {\bibfnamefont {David}\ \bibnamefont
  {Delepine}}\ and\ \bibinfo {author} {\bibfnamefont {Francesco}\ \bibnamefont
  {Vissani}},\ }\bibfield  {title} {\enquote {\bibinfo {title} {{Indirect
  bounds on $Z\to \mu e$ and lepton flavor violation at future colliders}},}\
  }\href {\doibase 10.1016/S0370-2693(01)01254-0} {\bibfield  {journal}
  {\bibinfo  {journal} {Phys. Lett. B}\ }\textbf {\bibinfo {volume} {522}},\
  \bibinfo {pages} {95--101} (\bibinfo {year} {2001})},\ \Eprint
  {http://arxiv.org/abs/hep-ph/0106287} {arXiv:hep-ph/0106287} \BibitemShut
  {NoStop}%
\bibitem [{\citenamefont {Gutsche}\ \emph {et~al.}(2011)\citenamefont
  {Gutsche}, \citenamefont {Helo}, \citenamefont {Kovalenko},\ and\
  \citenamefont {Lyubovitskij}}]{Gutsche:2011bi}%
  \BibitemOpen
  \bibfield  {author} {\bibinfo {author} {\bibfnamefont {Thomas}\ \bibnamefont
  {Gutsche}}, \bibinfo {author} {\bibfnamefont {Juan~C.}\ \bibnamefont {Helo}},
  \bibinfo {author} {\bibfnamefont {Sergey}\ \bibnamefont {Kovalenko}}, \ and\
  \bibinfo {author} {\bibfnamefont {Valery~E.}\ \bibnamefont {Lyubovitskij}},\
  }\bibfield  {title} {\enquote {\bibinfo {title} {{New bounds on lepton flavor
  violating decays of vector mesons and the $Z$ boson}},}\ }\href {\doibase
  10.1103/PhysRevD.83.115015} {\bibfield  {journal} {\bibinfo  {journal} {Phys.
  Rev. D}\ }\textbf {\bibinfo {volume} {83}},\ \bibinfo {pages} {115015}
  (\bibinfo {year} {2011})},\ \Eprint {http://arxiv.org/abs/1103.1317}
  {arXiv:1103.1317 [hep-ph]} \BibitemShut {NoStop}%
\bibitem [{\citenamefont {Davidson}\ \emph {et~al.}(2012)\citenamefont
  {Davidson}, \citenamefont {Lacroix},\ and\ \citenamefont
  {Verdier}}]{Davidson:2012wn}%
  \BibitemOpen
  \bibfield  {author} {\bibinfo {author} {\bibfnamefont {Sacha}\ \bibnamefont
  {Davidson}}, \bibinfo {author} {\bibfnamefont {Sylvain}\ \bibnamefont
  {Lacroix}}, \ and\ \bibinfo {author} {\bibfnamefont {Patrice}\ \bibnamefont
  {Verdier}},\ }\bibfield  {title} {\enquote {\bibinfo {title} {{LHC
  sensitivity to lepton flavour violating Z boson decays}},}\ }\href {\doibase
  10.1007/JHEP09(2012)092} {\bibfield  {journal} {\bibinfo  {journal} {JHEP}\
  }\textbf {\bibinfo {volume} {09}},\ \bibinfo {pages} {092} (\bibinfo {year}
  {2012})},\ \Eprint {http://arxiv.org/abs/1207.4894} {arXiv:1207.4894
  [hep-ph]} \BibitemShut {NoStop}%
\bibitem [{\citenamefont {Calibbi}\ \emph {et~al.}(2021)\citenamefont
  {Calibbi}, \citenamefont {Marcano},\ and\ \citenamefont
  {Roy}}]{Calibbi:2021pyh}%
  \BibitemOpen
  \bibfield  {author} {\bibinfo {author} {\bibfnamefont {Lorenzo}\ \bibnamefont
  {Calibbi}}, \bibinfo {author} {\bibfnamefont {Xabier}\ \bibnamefont
  {Marcano}}, \ and\ \bibinfo {author} {\bibfnamefont {Joydeep}\ \bibnamefont
  {Roy}},\ }\bibfield  {title} {\enquote {\bibinfo {title} {{Z lepton flavour
  violation as a probe for new physics at future $e^+e^-$ colliders}},}\ }\href
  {\doibase 10.1140/epjc/s10052-021-09777-3} {\bibfield  {journal} {\bibinfo
  {journal} {Eur. Phys. J. C}\ }\textbf {\bibinfo {volume} {81}},\ \bibinfo
  {pages} {1054} (\bibinfo {year} {2021})},\ \Eprint
  {http://arxiv.org/abs/2107.10273} {arXiv:2107.10273 [hep-ph]} \BibitemShut
  {NoStop}%
\bibitem [{\citenamefont {Davidson}\ and\ \citenamefont
  {Grenier}(2010)}]{Davidson:2010xv}%
  \BibitemOpen
  \bibfield  {author} {\bibinfo {author} {\bibfnamefont {Sacha}\ \bibnamefont
  {Davidson}}\ and\ \bibinfo {author} {\bibfnamefont {Gerald~Jean}\
  \bibnamefont {Grenier}},\ }\bibfield  {title} {\enquote {\bibinfo {title}
  {{Lepton flavour violating Higgs and tau to mu gamma}},}\ }\href {\doibase
  10.1103/PhysRevD.81.095016} {\bibfield  {journal} {\bibinfo  {journal} {Phys.
  Rev. D}\ }\textbf {\bibinfo {volume} {81}},\ \bibinfo {pages} {095016}
  (\bibinfo {year} {2010})},\ \Eprint {http://arxiv.org/abs/1001.0434}
  {arXiv:1001.0434 [hep-ph]} \BibitemShut {NoStop}%
\bibitem [{\citenamefont {Blankenburg}\ \emph {et~al.}(2012)\citenamefont
  {Blankenburg}, \citenamefont {Ellis},\ and\ \citenamefont
  {Isidori}}]{Blankenburg:2012ex}%
  \BibitemOpen
  \bibfield  {author} {\bibinfo {author} {\bibfnamefont {Gianluca}\
  \bibnamefont {Blankenburg}}, \bibinfo {author} {\bibfnamefont {John}\
  \bibnamefont {Ellis}}, \ and\ \bibinfo {author} {\bibfnamefont {Gino}\
  \bibnamefont {Isidori}},\ }\bibfield  {title} {\enquote {\bibinfo {title}
  {{Flavour-Changing Decays of a 125 GeV Higgs-like Particle}},}\ }\href
  {\doibase 10.1016/j.physletb.2012.05.007} {\bibfield  {journal} {\bibinfo
  {journal} {Phys. Lett. B}\ }\textbf {\bibinfo {volume} {712}},\ \bibinfo
  {pages} {386--390} (\bibinfo {year} {2012})},\ \Eprint
  {http://arxiv.org/abs/1202.5704} {arXiv:1202.5704 [hep-ph]} \BibitemShut
  {NoStop}%
\bibitem [{\citenamefont {Harnik}\ \emph {et~al.}(2013)\citenamefont {Harnik},
  \citenamefont {Kopp},\ and\ \citenamefont {Zupan}}]{Harnik:2012pb}%
  \BibitemOpen
  \bibfield  {author} {\bibinfo {author} {\bibfnamefont {Roni}\ \bibnamefont
  {Harnik}}, \bibinfo {author} {\bibfnamefont {Joachim}\ \bibnamefont {Kopp}},
  \ and\ \bibinfo {author} {\bibfnamefont {Jure}\ \bibnamefont {Zupan}},\
  }\bibfield  {title} {\enquote {\bibinfo {title} {{Flavor Violating Higgs
  Decays}},}\ }\href {\doibase 10.1007/JHEP03(2013)026} {\bibfield  {journal}
  {\bibinfo  {journal} {JHEP}\ }\textbf {\bibinfo {volume} {03}},\ \bibinfo
  {pages} {026} (\bibinfo {year} {2013})},\ \Eprint
  {http://arxiv.org/abs/1209.1397} {arXiv:1209.1397 [hep-ph]} \BibitemShut
  {NoStop}%
\bibitem [{\citenamefont {Altmannshofer}\ \emph {et~al.}(2016)\citenamefont
  {Altmannshofer}, \citenamefont {Gori}, \citenamefont {Kagan}, \citenamefont
  {Silvestrini},\ and\ \citenamefont {Zupan}}]{Altmannshofer:2015esa}%
  \BibitemOpen
  \bibfield  {author} {\bibinfo {author} {\bibfnamefont {Wolfgang}\
  \bibnamefont {Altmannshofer}}, \bibinfo {author} {\bibfnamefont {Stefania}\
  \bibnamefont {Gori}}, \bibinfo {author} {\bibfnamefont {Alexander~L.}\
  \bibnamefont {Kagan}}, \bibinfo {author} {\bibfnamefont {Luca}\ \bibnamefont
  {Silvestrini}}, \ and\ \bibinfo {author} {\bibfnamefont {Jure}\ \bibnamefont
  {Zupan}},\ }\bibfield  {title} {\enquote {\bibinfo {title} {{Uncovering Mass
  Generation Through Higgs Flavor Violation}},}\ }\href {\doibase
  10.1103/PhysRevD.93.031301} {\bibfield  {journal} {\bibinfo  {journal} {Phys.
  Rev. D}\ }\textbf {\bibinfo {volume} {93}},\ \bibinfo {pages} {031301}
  (\bibinfo {year} {2016})},\ \Eprint {http://arxiv.org/abs/1507.07927}
  {arXiv:1507.07927 [hep-ph]} \BibitemShut {NoStop}%
\bibitem [{\citenamefont {Davidson}\ \emph {et~al.}(2015)\citenamefont
  {Davidson}, \citenamefont {Mangano}, \citenamefont {Perries},\ and\
  \citenamefont {Sordini}}]{Davidson:2015zza}%
  \BibitemOpen
  \bibfield  {author} {\bibinfo {author} {\bibfnamefont {Sacha}\ \bibnamefont
  {Davidson}}, \bibinfo {author} {\bibfnamefont {Michelangelo~L.}\ \bibnamefont
  {Mangano}}, \bibinfo {author} {\bibfnamefont {Stephane}\ \bibnamefont
  {Perries}}, \ and\ \bibinfo {author} {\bibfnamefont {Viola}\ \bibnamefont
  {Sordini}},\ }\bibfield  {title} {\enquote {\bibinfo {title} {{Lepton Flavour
  Violating top decays at the LHC}},}\ }\href {\doibase
  10.1140/epjc/s10052-015-3649-5} {\bibfield  {journal} {\bibinfo  {journal}
  {Eur. Phys. J. C}\ }\textbf {\bibinfo {volume} {75}},\ \bibinfo {pages} {450}
  (\bibinfo {year} {2015})},\ \Eprint {http://arxiv.org/abs/1507.07163}
  {arXiv:1507.07163 [hep-ph]} \BibitemShut {NoStop}%
\bibitem [{\citenamefont {Altmannshofer}\ \emph {et~al.}(2025)\citenamefont
  {Altmannshofer}, \citenamefont {Balme}, \citenamefont {Donohue},
  \citenamefont {Gori},\ and\ \citenamefont
  {Mukundhan}}]{Altmannshofer:2025lun}%
  \BibitemOpen
  \bibfield  {author} {\bibinfo {author} {\bibfnamefont {Wolfgang}\
  \bibnamefont {Altmannshofer}}, \bibinfo {author} {\bibfnamefont {Zev}\
  \bibnamefont {Balme}}, \bibinfo {author} {\bibfnamefont {Christopher~M.}\
  \bibnamefont {Donohue}}, \bibinfo {author} {\bibfnamefont {Stefania}\
  \bibnamefont {Gori}}, \ and\ \bibinfo {author} {\bibfnamefont
  {Siddharth~Vignesh}\ \bibnamefont {Mukundhan}},\ }\bibfield  {title}
  {\enquote {\bibinfo {title} {{Targets for Flavor-Violating Top Decay}},}\
  }\href@noop {} {\  (\bibinfo {year} {2025})},\ \Eprint
  {http://arxiv.org/abs/2504.18664} {arXiv:2504.18664 [hep-ph]} \BibitemShut
  {NoStop}%
\bibitem [{\citenamefont {Altmannshofer}\ \emph {et~al.}(2022)\citenamefont
  {Altmannshofer}, \citenamefont {Caillol}, \citenamefont {Dam}, \citenamefont
  {Xella},\ and\ \citenamefont {Zhang}}]{Altmannshofer:2022fvz}%
  \BibitemOpen
  \bibfield  {author} {\bibinfo {author} {\bibfnamefont {Wolfgang}\
  \bibnamefont {Altmannshofer}}, \bibinfo {author} {\bibfnamefont {Cecile}\
  \bibnamefont {Caillol}}, \bibinfo {author} {\bibfnamefont {Mogens}\
  \bibnamefont {Dam}}, \bibinfo {author} {\bibfnamefont {Stefania}\
  \bibnamefont {Xella}}, \ and\ \bibinfo {author} {\bibfnamefont {Yongchao}\
  \bibnamefont {Zhang}},\ }\bibfield  {title} {\enquote {\bibinfo {title}
  {{Charged Lepton Flavour Violation in Heavy Particle DEcays}},}\ }in\
  \href@noop {} {\emph {\bibinfo {booktitle} {{Snowmass 2021}}}}\ (\bibinfo
  {year} {2022})\ \Eprint {http://arxiv.org/abs/2205.10576} {arXiv:2205.10576
  [hep-ph]} \BibitemShut {NoStop}%
\bibitem [{\citenamefont {Grzadkowski}\ \emph {et~al.}(2010)\citenamefont
  {Grzadkowski}, \citenamefont {Iskrzynski}, \citenamefont {Misiak},\ and\
  \citenamefont {Rosiek}}]{Grzadkowski:2010es}%
  \BibitemOpen
  \bibfield  {author} {\bibinfo {author} {\bibfnamefont {B.}~\bibnamefont
  {Grzadkowski}}, \bibinfo {author} {\bibfnamefont {M.}~\bibnamefont
  {Iskrzynski}}, \bibinfo {author} {\bibfnamefont {M.}~\bibnamefont {Misiak}},
  \ and\ \bibinfo {author} {\bibfnamefont {J.}~\bibnamefont {Rosiek}},\
  }\bibfield  {title} {\enquote {\bibinfo {title} {{Dimension-Six Terms in the
  Standard Model Lagrangian}},}\ }\href {\doibase 10.1007/JHEP10(2010)085}
  {\bibfield  {journal} {\bibinfo  {journal} {JHEP}\ }\textbf {\bibinfo
  {volume} {10}},\ \bibinfo {pages} {085} (\bibinfo {year} {2010})},\ \Eprint
  {http://arxiv.org/abs/1008.4884} {arXiv:1008.4884 [hep-ph]} \BibitemShut
  {NoStop}%
\bibitem [{\citenamefont {Altmannshofer}\ \emph {et~al.}(2023)\citenamefont
  {Altmannshofer}, \citenamefont {Munbodh},\ and\ \citenamefont
  {Oh}}]{Altmannshofer:2023tsa}%
  \BibitemOpen
  \bibfield  {author} {\bibinfo {author} {\bibfnamefont {Wolfgang}\
  \bibnamefont {Altmannshofer}}, \bibinfo {author} {\bibfnamefont {Pankaj}\
  \bibnamefont {Munbodh}}, \ and\ \bibinfo {author} {\bibfnamefont {Talise}\
  \bibnamefont {Oh}},\ }\bibfield  {title} {\enquote {\bibinfo {title}
  {{Probing lepton flavor violation at Circular Electron-Positron
  Colliders}},}\ }\href {\doibase 10.1007/JHEP08(2023)026} {\bibfield
  {journal} {\bibinfo  {journal} {JHEP}\ }\textbf {\bibinfo {volume} {08}},\
  \bibinfo {pages} {026} (\bibinfo {year} {2023})},\ \Eprint
  {http://arxiv.org/abs/2305.03869} {arXiv:2305.03869 [hep-ph]} \BibitemShut
  {NoStop}%
\bibitem [{\citenamefont {Munbodh}(2024)}]{Munbodh:2024shg}%
  \BibitemOpen
  \bibfield  {author} {\bibinfo {author} {\bibfnamefont {Pankaj}\ \bibnamefont
  {Munbodh}},\ }\bibfield  {title} {\enquote {\bibinfo {title} {{Lepton Flavor
  Violation at FCC-ee and CEPC}},}\ }in\ \href@noop {} {\emph {\bibinfo
  {booktitle} {{17th International Workshop on Tau Lepton Physics}}}}\
  (\bibinfo {year} {2024})\ \Eprint {http://arxiv.org/abs/2406.01935}
  {arXiv:2406.01935 [hep-ph]} \BibitemShut {NoStop}%
\bibitem [{\citenamefont {Bernardi}\ \emph {et~al.}(2022)\citenamefont
  {Bernardi} \emph {et~al.}}]{Bernardi:2022hny}%
  \BibitemOpen
  \bibfield  {author} {\bibinfo {author} {\bibfnamefont {G.}~\bibnamefont
  {Bernardi}} \emph {et~al.},\ }\bibfield  {title} {\enquote {\bibinfo {title}
  {{The Future Circular Collider: a Summary for the US 2021 Snowmass
  Process}},}\ }\href@noop {} {\  (\bibinfo {year} {2022})},\ \Eprint
  {http://arxiv.org/abs/2203.06520} {arXiv:2203.06520 [hep-ex]} \BibitemShut
  {NoStop}%
\bibitem [{\citenamefont {Benedikt}\ \emph {et~al.}(2025)\citenamefont
  {Benedikt} \emph {et~al.}}]{FCC:2025lpp}%
  \BibitemOpen
  \bibfield  {author} {\bibinfo {author} {\bibfnamefont {M.}~\bibnamefont
  {Benedikt}} \emph {et~al.} (\bibinfo {collaboration} {FCC}),\ }\bibfield
  {title} {\enquote {\bibinfo {title} {{Future Circular Collider Feasibility
  Study Report: Volume 1, Physics, Experiments, Detectors}},}\ }\href {\doibase
  10.17181/CERN.9DKX.TDH9} {\  (\bibinfo {year} {2025}),\
  10.17181/CERN.9DKX.TDH9},\ \Eprint {http://arxiv.org/abs/2505.00272}
  {arXiv:2505.00272 [hep-ex]} \BibitemShut {NoStop}%
\bibitem [{\citenamefont {Dong}\ \emph {et~al.}(2018)\citenamefont {Dong} \emph
  {et~al.}}]{CEPCStudyGroup:2018ghi}%
  \BibitemOpen
  \bibfield  {author} {\bibinfo {author} {\bibfnamefont {Mingyi}\ \bibnamefont
  {Dong}} \emph {et~al.} (\bibinfo {collaboration} {CEPC Study Group}),\
  }\bibfield  {title} {\enquote {\bibinfo {title} {{CEPC Conceptual Design
  Report: Volume 2 - Physics \& Detector}},}\ }\href@noop {} {\  (\bibinfo
  {year} {2018})},\ \Eprint {http://arxiv.org/abs/1811.10545} {arXiv:1811.10545
  [hep-ex]} \BibitemShut {NoStop}%
\bibitem [{\citenamefont {Abdallah}\ \emph {et~al.}(2024)\citenamefont
  {Abdallah} \emph {et~al.}}]{CEPCStudyGroup:2023quu}%
  \BibitemOpen
  \bibfield  {author} {\bibinfo {author} {\bibfnamefont {Waleed}\ \bibnamefont
  {Abdallah}} \emph {et~al.} (\bibinfo {collaboration} {CEPC Study Group}),\
  }\bibfield  {title} {\enquote {\bibinfo {title} {{CEPC Technical Design
  Report: Accelerator}},}\ }\href {\doibase 10.1007/s41605-024-00463-y}
  {\bibfield  {journal} {\bibinfo  {journal} {Radiat. Detect. Technol.
  Methods}\ }\textbf {\bibinfo {volume} {8}},\ \bibinfo {pages} {1--1105}
  (\bibinfo {year} {2024})},\ \bibinfo {note} {[Erratum:
  Radiat.Detect.Technol.Methods 9, 184--192 (2025)]},\ \Eprint
  {http://arxiv.org/abs/2312.14363} {arXiv:2312.14363 [physics.acc-ph]}
  \BibitemShut {NoStop}%
\bibitem [{\citenamefont {Ai}\ \emph {et~al.}(2024)\citenamefont {Ai} \emph
  {et~al.}}]{Ai:2024nmn}%
  \BibitemOpen
  \bibfield  {author} {\bibinfo {author} {\bibfnamefont {Xiaocong}\
  \bibnamefont {Ai}} \emph {et~al.},\ }\bibfield  {title} {\enquote {\bibinfo
  {title} {{Flavor Physics at CEPC: a General Perspective}},}\ }\href@noop {}
  {\  (\bibinfo {year} {2024})},\ \Eprint {http://arxiv.org/abs/2412.19743}
  {arXiv:2412.19743 [hep-ex]} \BibitemShut {NoStop}%
\bibitem [{\citenamefont {Han}\ \emph {et~al.}(2010)\citenamefont {Han},
  \citenamefont {Lewis},\ and\ \citenamefont {Sher}}]{Han:2010sa}%
  \BibitemOpen
  \bibfield  {author} {\bibinfo {author} {\bibfnamefont {Tao}\ \bibnamefont
  {Han}}, \bibinfo {author} {\bibfnamefont {Ian}\ \bibnamefont {Lewis}}, \ and\
  \bibinfo {author} {\bibfnamefont {Marc}\ \bibnamefont {Sher}},\ }\bibfield
  {title} {\enquote {\bibinfo {title} {{Mu-Tau Production at Hadron
  Colliders}},}\ }\href {\doibase 10.1007/JHEP03(2010)090} {\bibfield
  {journal} {\bibinfo  {journal} {JHEP}\ }\textbf {\bibinfo {volume} {03}},\
  \bibinfo {pages} {090} (\bibinfo {year} {2010})},\ \Eprint
  {http://arxiv.org/abs/1001.0022} {arXiv:1001.0022 [hep-ph]} \BibitemShut
  {NoStop}%
\bibitem [{\citenamefont {Murakami}\ and\ \citenamefont
  {Tait}(2015)}]{Murakami:2014tna}%
  \BibitemOpen
  \bibfield  {author} {\bibinfo {author} {\bibfnamefont {Brandon}\ \bibnamefont
  {Murakami}}\ and\ \bibinfo {author} {\bibfnamefont {Tim M.~P.}\ \bibnamefont
  {Tait}},\ }\bibfield  {title} {\enquote {\bibinfo {title} {{Searching for
  lepton flavor violation at a future high energy e+e- collider}},}\ }\href
  {\doibase 10.1103/PhysRevD.91.015002} {\bibfield  {journal} {\bibinfo
  {journal} {Phys. Rev. D}\ }\textbf {\bibinfo {volume} {91}},\ \bibinfo
  {pages} {015002} (\bibinfo {year} {2015})},\ \Eprint
  {http://arxiv.org/abs/1410.1485} {arXiv:1410.1485 [hep-ph]} \BibitemShut
  {NoStop}%
\bibitem [{\citenamefont {Cho}\ \emph {et~al.}(2019)\citenamefont {Cho},
  \citenamefont {Fukuda},\ and\ \citenamefont {Kono}}]{Cho:2018mro}%
  \BibitemOpen
  \bibfield  {author} {\bibinfo {author} {\bibfnamefont {Gi-Chol}\ \bibnamefont
  {Cho}}, \bibinfo {author} {\bibfnamefont {Yuka}\ \bibnamefont {Fukuda}}, \
  and\ \bibinfo {author} {\bibfnamefont {Takanori}\ \bibnamefont {Kono}},\
  }\bibfield  {title} {\enquote {\bibinfo {title} {{Lepton flavor violation via
  four-Fermi contact interactions at the International Linear Collider}},}\
  }\href {\doibase 10.1016/j.physletb.2018.12.056} {\bibfield  {journal}
  {\bibinfo  {journal} {Phys. Lett. B}\ }\textbf {\bibinfo {volume} {789}},\
  \bibinfo {pages} {399--404} (\bibinfo {year} {2019})},\ \Eprint
  {http://arxiv.org/abs/1803.10475} {arXiv:1803.10475 [hep-ph]} \BibitemShut
  {NoStop}%
\bibitem [{\citenamefont {Cirigliano}\ \emph {et~al.}(2021)\citenamefont
  {Cirigliano}, \citenamefont {Fuyuto}, \citenamefont {Lee}, \citenamefont
  {Mereghetti},\ and\ \citenamefont {Yan}}]{Cirigliano:2021img}%
  \BibitemOpen
  \bibfield  {author} {\bibinfo {author} {\bibfnamefont {Vincenzo}\
  \bibnamefont {Cirigliano}}, \bibinfo {author} {\bibfnamefont {Kaori}\
  \bibnamefont {Fuyuto}}, \bibinfo {author} {\bibfnamefont {Christopher}\
  \bibnamefont {Lee}}, \bibinfo {author} {\bibfnamefont {Emanuele}\
  \bibnamefont {Mereghetti}}, \ and\ \bibinfo {author} {\bibfnamefont {Bin}\
  \bibnamefont {Yan}},\ }\bibfield  {title} {\enquote {\bibinfo {title}
  {{Charged Lepton Flavor Violation at the EIC}},}\ }\href {\doibase
  10.1007/JHEP03(2021)256} {\bibfield  {journal} {\bibinfo  {journal} {JHEP}\
  }\textbf {\bibinfo {volume} {03}},\ \bibinfo {pages} {256} (\bibinfo {year}
  {2021})},\ \Eprint {http://arxiv.org/abs/2102.06176} {arXiv:2102.06176
  [hep-ph]} \BibitemShut {NoStop}%
\bibitem [{\citenamefont {Etesami}\ \emph {et~al.}(2021)\citenamefont
  {Etesami}, \citenamefont {Jafari}, \citenamefont {Najafabadi},\ and\
  \citenamefont {Tizchang}}]{Etesami:2021hex}%
  \BibitemOpen
  \bibfield  {author} {\bibinfo {author} {\bibfnamefont {S.~M.}\ \bibnamefont
  {Etesami}}, \bibinfo {author} {\bibfnamefont {R.}~\bibnamefont {Jafari}},
  \bibinfo {author} {\bibfnamefont {M.~Mohammadi}\ \bibnamefont {Najafabadi}},
  \ and\ \bibinfo {author} {\bibfnamefont {S.}~\bibnamefont {Tizchang}},\
  }\bibfield  {title} {\enquote {\bibinfo {title} {{Searching for lepton flavor
  violating interactions at future electron-positron colliders}},}\ }\href
  {\doibase 10.1103/PhysRevD.104.015034} {\bibfield  {journal} {\bibinfo
  {journal} {Phys. Rev. D}\ }\textbf {\bibinfo {volume} {104}},\ \bibinfo
  {pages} {015034} (\bibinfo {year} {2021})},\ \Eprint
  {http://arxiv.org/abs/2107.00545} {arXiv:2107.00545 [hep-ph]} \BibitemShut
  {NoStop}%
\bibitem [{\citenamefont {Calibbi}\ \emph {et~al.}(2022)\citenamefont
  {Calibbi}, \citenamefont {Li}, \citenamefont {Marcano},\ and\ \citenamefont
  {Schmidt}}]{Calibbi:2022ddo}%
  \BibitemOpen
  \bibfield  {author} {\bibinfo {author} {\bibfnamefont {Lorenzo}\ \bibnamefont
  {Calibbi}}, \bibinfo {author} {\bibfnamefont {Tong}\ \bibnamefont {Li}},
  \bibinfo {author} {\bibfnamefont {Xabier}\ \bibnamefont {Marcano}}, \ and\
  \bibinfo {author} {\bibfnamefont {Michael~A.}\ \bibnamefont {Schmidt}},\
  }\bibfield  {title} {\enquote {\bibinfo {title} {{Indirect constraints on
  lepton-flavor-violating quarkonium decays}},}\ }\href {\doibase
  10.1103/PhysRevD.106.115039} {\bibfield  {journal} {\bibinfo  {journal}
  {Phys. Rev. D}\ }\textbf {\bibinfo {volume} {106}},\ \bibinfo {pages}
  {115039} (\bibinfo {year} {2022})},\ \Eprint
  {http://arxiv.org/abs/2207.10913} {arXiv:2207.10913 [hep-ph]} \BibitemShut
  {NoStop}%
\bibitem [{\citenamefont {Barik}\ \emph {et~al.}(2023)\citenamefont {Barik},
  \citenamefont {Dey},\ and\ \citenamefont {Samui}}]{Barik:2023bgx}%
  \BibitemOpen
  \bibfield  {author} {\bibinfo {author} {\bibfnamefont {Anjan~Kumar}\
  \bibnamefont {Barik}}, \bibinfo {author} {\bibfnamefont {Atri}\ \bibnamefont
  {Dey}}, \ and\ \bibinfo {author} {\bibfnamefont {Tousik}\ \bibnamefont
  {Samui}},\ }\bibfield  {title} {\enquote {\bibinfo {title} {{Search for
  Lepton Flavour Violating Signals at the Future Electron-Proton Colliders}},}\
  }\href@noop {} {\  (\bibinfo {year} {2023})},\ \Eprint
  {http://arxiv.org/abs/2306.10540} {arXiv:2306.10540 [hep-ph]} \BibitemShut
  {NoStop}%
\bibitem [{\citenamefont {Lichtenstein}\ \emph {et~al.}(2023)\citenamefont
  {Lichtenstein}, \citenamefont {Schmidt}, \citenamefont {Valencia},\ and\
  \citenamefont {Volkas}}]{Lichtenstein:2023iut}%
  \BibitemOpen
  \bibfield  {author} {\bibinfo {author} {\bibfnamefont {Gabriela}\
  \bibnamefont {Lichtenstein}}, \bibinfo {author} {\bibfnamefont {Michael~A.}\
  \bibnamefont {Schmidt}}, \bibinfo {author} {\bibfnamefont {German}\
  \bibnamefont {Valencia}}, \ and\ \bibinfo {author} {\bibfnamefont
  {Raymond~R.}\ \bibnamefont {Volkas}},\ }\bibfield  {title} {\enquote
  {\bibinfo {title} {{Complementarity of $\mu$TRISTAN and Belle II in searches
  for charged-lepton flavour violation}},}\ }\href {\doibase
  10.1016/j.physletb.2023.138144} {\bibfield  {journal} {\bibinfo  {journal}
  {Phys. Lett. B}\ }\textbf {\bibinfo {volume} {845}},\ \bibinfo {pages}
  {138144} (\bibinfo {year} {2023})},\ \Eprint
  {http://arxiv.org/abs/2307.11369} {arXiv:2307.11369 [hep-ph]} \BibitemShut
  {NoStop}%
\bibitem [{\citenamefont {Liu}\ \emph {et~al.}(2024)\citenamefont {Liu},
  \citenamefont {Song},\ and\ \citenamefont {Zhang}}]{Liu:2024gui}%
  \BibitemOpen
  \bibfield  {author} {\bibinfo {author} {\bibfnamefont {Jia}\ \bibnamefont
  {Liu}}, \bibinfo {author} {\bibfnamefont {Muyuan}\ \bibnamefont {Song}}, \
  and\ \bibinfo {author} {\bibfnamefont {Haohao}\ \bibnamefont {Zhang}},\
  }\bibfield  {title} {\enquote {\bibinfo {title} {{Revisiting for maximal
  flavor violating $ {Z}_{e\mu}^{\prime } $ and its phenomenology
  constraints}},}\ }\href {\doibase 10.1007/JHEP10(2024)128} {\bibfield
  {journal} {\bibinfo  {journal} {JHEP}\ }\textbf {\bibinfo {volume} {10}},\
  \bibinfo {pages} {128} (\bibinfo {year} {2024})},\ \Eprint
  {http://arxiv.org/abs/2407.05831} {arXiv:2407.05831 [hep-ph]} \BibitemShut
  {NoStop}%
\bibitem [{\citenamefont {Jahedi}\ and\ \citenamefont
  {Sarkar}(2024)}]{Jahedi:2024kvi}%
  \BibitemOpen
  \bibfield  {author} {\bibinfo {author} {\bibfnamefont {Sahabub}\ \bibnamefont
  {Jahedi}}\ and\ \bibinfo {author} {\bibfnamefont {Abhik}\ \bibnamefont
  {Sarkar}},\ }\bibfield  {title} {\enquote {\bibinfo {title} {{Exploring
  optimal sensitivity of lepton flavor violating effective couplings at the
  e+e- colliders}},}\ }\href {\doibase 10.1103/PhysRevD.110.095021} {\bibfield
  {journal} {\bibinfo  {journal} {Phys. Rev. D}\ }\textbf {\bibinfo {volume}
  {110}},\ \bibinfo {pages} {095021} (\bibinfo {year} {2024})},\ \Eprint
  {http://arxiv.org/abs/2408.00190} {arXiv:2408.00190 [hep-ph]} \BibitemShut
  {NoStop}%
\bibitem [{\citenamefont {De}(2024)}]{De:2024foq}%
  \BibitemOpen
  \bibfield  {author} {\bibinfo {author} {\bibfnamefont {Bibhabasu}\
  \bibnamefont {De}},\ }\bibfield  {title} {\enquote {\bibinfo {title}
  {{Leptoquark-induced CLFV decays with a light SM-singlet scalar}},}\ }\href
  {\doibase 10.1016/j.physletb.2024.138784} {\bibfield  {journal} {\bibinfo
  {journal} {Phys. Lett. B}\ }\textbf {\bibinfo {volume} {855}},\ \bibinfo
  {pages} {138784} (\bibinfo {year} {2024})},\ \Eprint
  {http://arxiv.org/abs/2405.06970} {arXiv:2405.06970 [hep-ph]} \BibitemShut
  {NoStop}%
\bibitem [{ILC(2013)}]{ILC:2013jhg}%
  \BibitemOpen
  \bibfield  {title} {\enquote {\bibinfo {title} {{The International Linear
  Collider Technical Design Report - Volume 2: Physics}},}\ }\href@noop {} {\
  (\bibinfo {year} {2013})},\ \Eprint {http://arxiv.org/abs/1306.6352}
  {arXiv:1306.6352 [hep-ph]} \BibitemShut {NoStop}%
\bibitem [{\citenamefont {Bambade}\ \emph {et~al.}(2019)\citenamefont {Bambade}
  \emph {et~al.}}]{Bambade:2019fyw}%
  \BibitemOpen
  \bibfield  {author} {\bibinfo {author} {\bibfnamefont {Philip}\ \bibnamefont
  {Bambade}} \emph {et~al.},\ }\bibfield  {title} {\enquote {\bibinfo {title}
  {{The International Linear Collider: A Global Project}},}\ }\href@noop {} {\
  (\bibinfo {year} {2019})},\ \Eprint {http://arxiv.org/abs/1903.01629}
  {arXiv:1903.01629 [hep-ex]} \BibitemShut {NoStop}%
\bibitem [{\citenamefont {Aryshev}\ \emph {et~al.}(2022)\citenamefont {Aryshev}
  \emph {et~al.}}]{ILCInternationalDevelopmentTeam:2022izu}%
  \BibitemOpen
  \bibfield  {author} {\bibinfo {author} {\bibfnamefont {Alexander}\
  \bibnamefont {Aryshev}} \emph {et~al.} (\bibinfo {collaboration} {ILC
  International Development Team}),\ }\bibfield  {title} {\enquote {\bibinfo
  {title} {{The International Linear Collider: Report to Snowmass 2021}},}\
  }\href@noop {} {\  (\bibinfo {year} {2022})},\ \Eprint
  {http://arxiv.org/abs/2203.07622} {arXiv:2203.07622 [physics.acc-ph]}
  \BibitemShut {NoStop}%
\bibitem [{Lin(2012)}]{Linssen:2012hp}%
  \BibitemOpen
  \bibfield  {title} {\enquote {\bibinfo {title} {{Physics and Detectors at
  CLIC: CLIC Conceptual Design Report}},}\ }\href {\doibase
  10.5170/CERN-2012-003} {\  (\bibinfo {year} {2012}),\
  10.5170/CERN-2012-003},\ \Eprint {http://arxiv.org/abs/1202.5940}
  {arXiv:1202.5940 [physics.ins-det]} \BibitemShut {NoStop}%
\bibitem [{\citenamefont {de~Blas}\ \emph {et~al.}(2018)\citenamefont {de~Blas}
  \emph {et~al.}}]{CLIC:2018fvx}%
  \BibitemOpen
  \bibfield  {author} {\bibinfo {author} {\bibfnamefont {J.}~\bibnamefont
  {de~Blas}} \emph {et~al.} (\bibinfo {collaboration} {CLIC}),\ }\bibfield
  {title} {\enquote {\bibinfo {title} {{The CLIC Potential for New Physics}},}\
  }\href {\doibase 10.23731/CYRM-2018-003} {\ \textbf {\bibinfo {volume}
  {3/2018}} (\bibinfo {year} {2018}),\ 10.23731/CYRM-2018-003},\ \Eprint
  {http://arxiv.org/abs/1812.02093} {arXiv:1812.02093 [hep-ph]} \BibitemShut
  {NoStop}%
\bibitem [{\citenamefont {Brunner}\ \emph {et~al.}(2022)\citenamefont {Brunner}
  \emph {et~al.}}]{Brunner:2022usy}%
  \BibitemOpen
  \bibfield  {author} {\bibinfo {author} {\bibfnamefont {O.}~\bibnamefont
  {Brunner}} \emph {et~al.},\ }\bibfield  {title} {\enquote {\bibinfo {title}
  {{The CLIC project}},}\ }\href@noop {} {\  (\bibinfo {year} {2022})},\
  \Eprint {http://arxiv.org/abs/2203.09186} {arXiv:2203.09186 [physics.acc-ph]}
  \BibitemShut {NoStop}%
\bibitem [{\citenamefont {Crivellin}\ \emph {et~al.}(2014)\citenamefont
  {Crivellin}, \citenamefont {Najjari},\ and\ \citenamefont
  {Rosiek}}]{Crivellin:2013hpa}%
  \BibitemOpen
  \bibfield  {author} {\bibinfo {author} {\bibfnamefont {Andreas}\ \bibnamefont
  {Crivellin}}, \bibinfo {author} {\bibfnamefont {Saereh}\ \bibnamefont
  {Najjari}}, \ and\ \bibinfo {author} {\bibfnamefont {Janusz}\ \bibnamefont
  {Rosiek}},\ }\bibfield  {title} {\enquote {\bibinfo {title} {{Lepton Flavor
  Violation in the Standard Model with general Dimension-Six Operators}},}\
  }\href {\doibase 10.1007/JHEP04(2014)167} {\bibfield  {journal} {\bibinfo
  {journal} {JHEP}\ }\textbf {\bibinfo {volume} {04}},\ \bibinfo {pages} {167}
  (\bibinfo {year} {2014})},\ \Eprint {http://arxiv.org/abs/1312.0634}
  {arXiv:1312.0634 [hep-ph]} \BibitemShut {NoStop}%
\bibitem [{\citenamefont {Celis}\ \emph {et~al.}(2014)\citenamefont {Celis},
  \citenamefont {Cirigliano},\ and\ \citenamefont {Passemar}}]{Celis:2014asa}%
  \BibitemOpen
  \bibfield  {author} {\bibinfo {author} {\bibfnamefont {Alejandro}\
  \bibnamefont {Celis}}, \bibinfo {author} {\bibfnamefont {Vincenzo}\
  \bibnamefont {Cirigliano}}, \ and\ \bibinfo {author} {\bibfnamefont {Emilie}\
  \bibnamefont {Passemar}},\ }\bibfield  {title} {\enquote {\bibinfo {title}
  {{Model-discriminating power of lepton flavor violating $\tau$ decays}},}\
  }\href {\doibase 10.1103/PhysRevD.89.095014} {\bibfield  {journal} {\bibinfo
  {journal} {Phys. Rev. D}\ }\textbf {\bibinfo {volume} {89}},\ \bibinfo
  {pages} {095014} (\bibinfo {year} {2014})},\ \Eprint
  {http://arxiv.org/abs/1403.5781} {arXiv:1403.5781 [hep-ph]} \BibitemShut
  {NoStop}%
\bibitem [{\citenamefont {Fern\'andez-Mart\'\i{}nez}\ \emph
  {et~al.}(2024)\citenamefont {Fern\'andez-Mart\'\i{}nez}, \citenamefont
  {Marcano},\ and\ \citenamefont {Naredo-Tuero}}]{Fernandez-Martinez:2024bxg}%
  \BibitemOpen
  \bibfield  {author} {\bibinfo {author} {\bibfnamefont {Enrique}\ \bibnamefont
  {Fern\'andez-Mart\'\i{}nez}}, \bibinfo {author} {\bibfnamefont {Xabier}\
  \bibnamefont {Marcano}}, \ and\ \bibinfo {author} {\bibfnamefont {Daniel}\
  \bibnamefont {Naredo-Tuero}},\ }\bibfield  {title} {\enquote {\bibinfo
  {title} {{Global lepton flavour violating constraints on new physics}},}\
  }\href {\doibase 10.1140/epjc/s10052-024-12973-6} {\bibfield  {journal}
  {\bibinfo  {journal} {Eur. Phys. J. C}\ }\textbf {\bibinfo {volume} {84}},\
  \bibinfo {pages} {666} (\bibinfo {year} {2024})},\ \Eprint
  {http://arxiv.org/abs/2403.09772} {arXiv:2403.09772 [hep-ph]} \BibitemShut
  {NoStop}%
\bibitem [{\citenamefont {Jenkins}\ \emph {et~al.}(2014)\citenamefont
  {Jenkins}, \citenamefont {Manohar},\ and\ \citenamefont
  {Trott}}]{Jenkins:2013wua}%
  \BibitemOpen
  \bibfield  {author} {\bibinfo {author} {\bibfnamefont {Elizabeth~E.}\
  \bibnamefont {Jenkins}}, \bibinfo {author} {\bibfnamefont {Aneesh~V.}\
  \bibnamefont {Manohar}}, \ and\ \bibinfo {author} {\bibfnamefont {Michael}\
  \bibnamefont {Trott}},\ }\bibfield  {title} {\enquote {\bibinfo {title}
  {{Renormalization Group Evolution of the Standard Model Dimension Six
  Operators II: Yukawa Dependence}},}\ }\href {\doibase
  10.1007/JHEP01(2014)035} {\bibfield  {journal} {\bibinfo  {journal} {JHEP}\
  }\textbf {\bibinfo {volume} {01}},\ \bibinfo {pages} {035} (\bibinfo {year}
  {2014})},\ \Eprint {http://arxiv.org/abs/1310.4838} {arXiv:1310.4838
  [hep-ph]} \BibitemShut {NoStop}%
\bibitem [{\citenamefont {Alonso}\ \emph {et~al.}(2014)\citenamefont {Alonso},
  \citenamefont {Jenkins}, \citenamefont {Manohar},\ and\ \citenamefont
  {Trott}}]{Alonso:2013hga}%
  \BibitemOpen
  \bibfield  {author} {\bibinfo {author} {\bibfnamefont {Rodrigo}\ \bibnamefont
  {Alonso}}, \bibinfo {author} {\bibfnamefont {Elizabeth~E.}\ \bibnamefont
  {Jenkins}}, \bibinfo {author} {\bibfnamefont {Aneesh~V.}\ \bibnamefont
  {Manohar}}, \ and\ \bibinfo {author} {\bibfnamefont {Michael}\ \bibnamefont
  {Trott}},\ }\bibfield  {title} {\enquote {\bibinfo {title} {{Renormalization
  Group Evolution of the Standard Model Dimension Six Operators III: Gauge
  Coupling Dependence and Phenomenology}},}\ }\href {\doibase
  10.1007/JHEP04(2014)159} {\bibfield  {journal} {\bibinfo  {journal} {JHEP}\
  }\textbf {\bibinfo {volume} {04}},\ \bibinfo {pages} {159} (\bibinfo {year}
  {2014})},\ \Eprint {http://arxiv.org/abs/1312.2014} {arXiv:1312.2014
  [hep-ph]} \BibitemShut {NoStop}%
\bibitem [{\citenamefont {Dam}(2019)}]{Dam:2018rfz}%
  \BibitemOpen
  \bibfield  {author} {\bibinfo {author} {\bibfnamefont {Mogens}\ \bibnamefont
  {Dam}},\ }\bibfield  {title} {\enquote {\bibinfo {title} {{Tau-lepton Physics
  at the FCC-ee circular e$^+$e$^-$ Collider}},}\ }\href {\doibase
  10.21468/SciPostPhysProc.1.041} {\bibfield  {journal} {\bibinfo  {journal}
  {SciPost Phys. Proc.}\ }\textbf {\bibinfo {volume} {1}},\ \bibinfo {pages}
  {041} (\bibinfo {year} {2019})},\ \Eprint {http://arxiv.org/abs/1811.09408}
  {arXiv:1811.09408 [hep-ex]} \BibitemShut {NoStop}%
\bibitem [{\citenamefont {Adolphsen}\ \emph {et~al.}(2013)\citenamefont
  {Adolphsen} \emph {et~al.}}]{Adolphsen:2013kya}%
  \BibitemOpen
  \bibfield  {author} {\bibinfo {author} {\bibfnamefont {Chris}\ \bibnamefont
  {Adolphsen}} \emph {et~al.},\ }\bibfield  {title} {\enquote {\bibinfo {title}
  {{The International Linear Collider Technical Design Report - Volume 3.II:
  Accelerator Baseline Design}},}\ }\href@noop {} {\  (\bibinfo {year}
  {2013})},\ \Eprint {http://arxiv.org/abs/1306.6328} {arXiv:1306.6328
  [physics.acc-ph]} \BibitemShut {NoStop}%
\bibitem [{\citenamefont {Charles}\ \emph {et~al.}(2018)\citenamefont {Charles}
  \emph {et~al.}}]{CLICdp:2018cto}%
  \BibitemOpen
  \bibfield  {author} {\bibinfo {author} {\bibfnamefont {T.~K.}\ \bibnamefont
  {Charles}} \emph {et~al.} (\bibinfo {collaboration} {CLICdp, CLIC}),\
  }\bibfield  {title} {\enquote {\bibinfo {title} {{The Compact Linear Collider
  (CLIC) - 2018 Summary Report}},}\ }\href {\doibase 10.23731/CYRM-2018-002} {\
  \textbf {\bibinfo {volume} {2/2018}} (\bibinfo {year} {2018}),\
  10.23731/CYRM-2018-002},\ \Eprint {http://arxiv.org/abs/1812.06018}
  {arXiv:1812.06018 [physics.acc-ph]} \BibitemShut {NoStop}%
\bibitem [{\citenamefont {Nicrosini}\ and\ \citenamefont
  {Trentadue}(1987)}]{NICROSINI1987551}%
  \BibitemOpen
  \bibfield  {author} {\bibinfo {author} {\bibfnamefont {O.}~\bibnamefont
  {Nicrosini}}\ and\ \bibinfo {author} {\bibfnamefont {Luca}\ \bibnamefont
  {Trentadue}},\ }\bibfield  {title} {\enquote {\bibinfo {title} {Soft photons
  and second order radiative corrections to $e^+e^- \to z^0$},}\ }\href
  {\doibase https://doi.org/10.1016/0370-2693(87)90819-7} {\bibfield  {journal}
  {\bibinfo  {journal} {Physics Letters B}\ }\textbf {\bibinfo {volume}
  {196}},\ \bibinfo {pages} {551--556} (\bibinfo {year} {1987})}\BibitemShut
  {NoStop}%
\bibitem [{\citenamefont {Garosi}\ \emph {et~al.}(2023)\citenamefont {Garosi},
  \citenamefont {Marzocca},\ and\ \citenamefont
  {Trifinopoulos}}]{Garosi:2023bvq}%
  \BibitemOpen
  \bibfield  {author} {\bibinfo {author} {\bibfnamefont {Francesco}\
  \bibnamefont {Garosi}}, \bibinfo {author} {\bibfnamefont {David}\
  \bibnamefont {Marzocca}}, \ and\ \bibinfo {author} {\bibfnamefont {Sokratis}\
  \bibnamefont {Trifinopoulos}},\ }\bibfield  {title} {\enquote {\bibinfo
  {title} {{LePDF: Standard Model PDFs for high-energy lepton colliders}},}\
  }\href {\doibase 10.1007/JHEP09(2023)107} {\bibfield  {journal} {\bibinfo
  {journal} {JHEP}\ }\textbf {\bibinfo {volume} {09}},\ \bibinfo {pages} {107}
  (\bibinfo {year} {2023})},\ \Eprint {http://arxiv.org/abs/2303.16964}
  {arXiv:2303.16964 [hep-ph]} \BibitemShut {NoStop}%
\bibitem [{\citenamefont {Kuraev}\ and\ \citenamefont
  {Fadin}(1985)}]{Kuraev:1985hb}%
  \BibitemOpen
  \bibfield  {author} {\bibinfo {author} {\bibfnamefont {E.~A.}\ \bibnamefont
  {Kuraev}}\ and\ \bibinfo {author} {\bibfnamefont {Victor~S.}\ \bibnamefont
  {Fadin}},\ }\bibfield  {title} {\enquote {\bibinfo {title} {{On Radiative
  Corrections to e+ e- Single Photon Annihilation at High-Energy}},}\
  }\href@noop {} {\bibfield  {journal} {\bibinfo  {journal} {Sov. J. Nucl.
  Phys.}\ }\textbf {\bibinfo {volume} {41}},\ \bibinfo {pages} {466--472}
  (\bibinfo {year} {1985})}\BibitemShut {NoStop}%
\bibitem [{\citenamefont {Jadach}\ \emph {et~al.}(2001)\citenamefont {Jadach},
  \citenamefont {Ward},\ and\ \citenamefont {Was}}]{Jadach:2000ir}%
  \BibitemOpen
  \bibfield  {author} {\bibinfo {author} {\bibfnamefont {S.}~\bibnamefont
  {Jadach}}, \bibinfo {author} {\bibfnamefont {B.~F.~L.}\ \bibnamefont {Ward}},
  \ and\ \bibinfo {author} {\bibfnamefont {Z.}~\bibnamefont {Was}},\ }\bibfield
   {title} {\enquote {\bibinfo {title} {{Coherent exclusive exponentiation for
  precision Monte Carlo calculations}},}\ }\href {\doibase
  10.1103/PhysRevD.63.113009} {\bibfield  {journal} {\bibinfo  {journal} {Phys.
  Rev. D}\ }\textbf {\bibinfo {volume} {63}},\ \bibinfo {pages} {113009}
  (\bibinfo {year} {2001})},\ \Eprint {http://arxiv.org/abs/hep-ph/0006359}
  {arXiv:hep-ph/0006359} \BibitemShut {NoStop}%
\bibitem [{\citenamefont {Greco}\ \emph {et~al.}(2016)\citenamefont {Greco},
  \citenamefont {Han},\ and\ \citenamefont {Liu}}]{Greco:2016izi}%
  \BibitemOpen
  \bibfield  {author} {\bibinfo {author} {\bibfnamefont {Mario}\ \bibnamefont
  {Greco}}, \bibinfo {author} {\bibfnamefont {Tao}\ \bibnamefont {Han}}, \ and\
  \bibinfo {author} {\bibfnamefont {Zhen}\ \bibnamefont {Liu}},\ }\bibfield
  {title} {\enquote {\bibinfo {title} {{ISR effects for resonant Higgs
  production at future lepton colliders}},}\ }\href {\doibase
  10.1016/j.physletb.2016.10.078} {\bibfield  {journal} {\bibinfo  {journal}
  {Phys. Lett. B}\ }\textbf {\bibinfo {volume} {763}},\ \bibinfo {pages}
  {409--415} (\bibinfo {year} {2016})},\ \Eprint
  {http://arxiv.org/abs/1607.03210} {arXiv:1607.03210 [hep-ph]} \BibitemShut
  {NoStop}%
\bibitem [{\citenamefont {Alwall}\ \emph {et~al.}(2014)\citenamefont {Alwall},
  \citenamefont {Frederix}, \citenamefont {Frixione}, \citenamefont {Hirschi},
  \citenamefont {Maltoni}, \citenamefont {Mattelaer}, \citenamefont {Shao},
  \citenamefont {Stelzer}, \citenamefont {Torrielli},\ and\ \citenamefont
  {Zaro}}]{Alwall:2014hca}%
  \BibitemOpen
  \bibfield  {author} {\bibinfo {author} {\bibfnamefont {J.}~\bibnamefont
  {Alwall}}, \bibinfo {author} {\bibfnamefont {R.}~\bibnamefont {Frederix}},
  \bibinfo {author} {\bibfnamefont {S.}~\bibnamefont {Frixione}}, \bibinfo
  {author} {\bibfnamefont {V.}~\bibnamefont {Hirschi}}, \bibinfo {author}
  {\bibfnamefont {F.}~\bibnamefont {Maltoni}}, \bibinfo {author} {\bibfnamefont
  {O.}~\bibnamefont {Mattelaer}}, \bibinfo {author} {\bibfnamefont {H.~S.}\
  \bibnamefont {Shao}}, \bibinfo {author} {\bibfnamefont {T.}~\bibnamefont
  {Stelzer}}, \bibinfo {author} {\bibfnamefont {P.}~\bibnamefont {Torrielli}},
  \ and\ \bibinfo {author} {\bibfnamefont {M.}~\bibnamefont {Zaro}},\
  }\bibfield  {title} {\enquote {\bibinfo {title} {{The automated computation
  of tree-level and next-to-leading order differential cross sections, and
  their matching to parton shower simulations}},}\ }\href {\doibase
  10.1007/JHEP07(2014)079} {\bibfield  {journal} {\bibinfo  {journal} {JHEP}\
  }\textbf {\bibinfo {volume} {07}},\ \bibinfo {pages} {079} (\bibinfo {year}
  {2014})},\ \Eprint {http://arxiv.org/abs/1405.0301} {arXiv:1405.0301
  [hep-ph]} \BibitemShut {NoStop}%
\bibitem [{\citenamefont {Navas}\ \emph {et~al.}(2024)\citenamefont {Navas}
  \emph {et~al.}}]{ParticleDataGroup:2024cfk}%
  \BibitemOpen
  \bibfield  {author} {\bibinfo {author} {\bibfnamefont {S.}~\bibnamefont
  {Navas}} \emph {et~al.} (\bibinfo {collaboration} {Particle Data Group}),\
  }\bibfield  {title} {\enquote {\bibinfo {title} {{Review of particle
  physics}},}\ }\href {\doibase 10.1103/PhysRevD.110.030001} {\bibfield
  {journal} {\bibinfo  {journal} {Phys. Rev. D}\ }\textbf {\bibinfo {volume}
  {110}},\ \bibinfo {pages} {030001} (\bibinfo {year} {2024})}\BibitemShut
  {NoStop}%
\bibitem [{\citenamefont {Behnke}\ \emph {et~al.}(2013)\citenamefont {Behnke}
  \emph {et~al.}}]{Behnke:2013lya}%
  \BibitemOpen
  \bibfield  {author} {\bibinfo {author} {\bibfnamefont {Ties}\ \bibnamefont
  {Behnke}} \emph {et~al.},\ }\bibfield  {title} {\enquote {\bibinfo {title}
  {{The International Linear Collider Technical Design Report - Volume 4:
  Detectors}},}\ }\href@noop {} {\  (\bibinfo {year} {2013})},\ \Eprint
  {http://arxiv.org/abs/1306.6329} {arXiv:1306.6329 [physics.ins-det]}
  \BibitemShut {NoStop}%
\bibitem [{\citenamefont {Balazs}\ \emph {et~al.}(2025)\citenamefont {Balazs}
  \emph {et~al.}}]{LinearCollider:2025lya}%
  \BibitemOpen
  \bibfield  {author} {\bibinfo {author} {\bibfnamefont {C.}~\bibnamefont
  {Balazs}} \emph {et~al.} (\bibinfo {collaboration} {Linear Collider}),\
  }\bibfield  {title} {\enquote {\bibinfo {title} {{The Linear Collider
  Facility (LCF) at CERN}},}\ }\href@noop {} {\  (\bibinfo {year} {2025})},\
  \Eprint {http://arxiv.org/abs/2503.24049} {arXiv:2503.24049 [hep-ex]}
  \BibitemShut {NoStop}%
\bibitem [{\citenamefont {Hayasaka}\ \emph {et~al.}(2010)\citenamefont
  {Hayasaka} \emph {et~al.}}]{Hayasaka:2010np}%
  \BibitemOpen
  \bibfield  {author} {\bibinfo {author} {\bibfnamefont {K.}~\bibnamefont
  {Hayasaka}} \emph {et~al.},\ }\bibfield  {title} {\enquote {\bibinfo {title}
  {{Search for Lepton Flavor Violating Tau Decays into Three Leptons with 719
  Million Produced Tau+Tau- Pairs}},}\ }\href {\doibase
  10.1016/j.physletb.2010.03.037} {\bibfield  {journal} {\bibinfo  {journal}
  {Phys. Lett. B}\ }\textbf {\bibinfo {volume} {687}},\ \bibinfo {pages}
  {139--143} (\bibinfo {year} {2010})},\ \Eprint
  {http://arxiv.org/abs/1001.3221} {arXiv:1001.3221 [hep-ex]} \BibitemShut
  {NoStop}%
\bibitem [{\citenamefont {Lees}\ \emph {et~al.}(2010)\citenamefont {Lees} \emph
  {et~al.}}]{BaBar:2010axs}%
  \BibitemOpen
  \bibfield  {author} {\bibinfo {author} {\bibfnamefont {J.~P.}\ \bibnamefont
  {Lees}} \emph {et~al.} (\bibinfo {collaboration} {BaBar}),\ }\bibfield
  {title} {\enquote {\bibinfo {title} {{Limits on tau Lepton-Flavor Violating
  Decays in three charged leptons}},}\ }\href {\doibase
  10.1103/PhysRevD.81.111101} {\bibfield  {journal} {\bibinfo  {journal} {Phys.
  Rev. D}\ }\textbf {\bibinfo {volume} {81}},\ \bibinfo {pages} {111101}
  (\bibinfo {year} {2010})},\ \Eprint {http://arxiv.org/abs/1002.4550}
  {arXiv:1002.4550 [hep-ex]} \BibitemShut {NoStop}%
\end{thebibliography}%

\end{document}